\documentclass[fleqn,usenatbib]{mnras}

\usepackage[T1]{fontenc}

\DeclareRobustCommand{\VAN}[3]{#2}
\let\VANthebibliography\thebibliography
\def\thebibliography{\DeclareRobustCommand{\VAN}[3]{##3}\VANthebibliography}

\usepackage{graphicx}	% Including figure files
\usepackage{amsmath}	% Advanced maths commands
\usepackage{amssymb}	% Extra maths symbols

\usepackage{newtxtext,newtxmath}

\title[Multi-Epoch Machine Learning]{Multi-Epoch Machine Learning 1: Unravelling Nature vs Nurture for Galaxy Formation}
\author[Robert McGibbon \& Sadegh Khochfar]{
Robert J. McGibbon,$^{1}$\thanks{E-mail: rob.mcgibbon@ed.ac.uk}
Sadegh Khochfar$^{1}$
\\
$^{1}$Institute for Astronomy, University of Edinburgh, Royal Observatory, Edinburgh EH9 3HJ
}

\date{Accepted 3 May 2022}
\pubyear{2022}

\begin{document}
\label{firstpage}
\pagerange{\pageref{firstpage}--\pageref{lastpage}}
\maketitle

\begin{abstract}
% This is a simple template for authors to write new MNRAS papers.
% The abstract should briefly describe the aims, methods, and main results of the paper.
% It should be a single paragraph not more than 250 words (200 words for Letters).
% No references should appear in the abstract.
We present a novel machine learning method for predicting the baryonic properties of dark matter only subhalos from N-body simulations. Our model is built using the extremely randomized tree (ERT) algorithm and takes subhalo properties over a wide range of redshifts as its input features.  We train our model using the IllustrisTNG simulations to predict blackhole mass, gas mass, magnitudes, star formation rate, stellar mass, and metallicity.  We compare the results of our method with a baseline model from previous works, and against a model that only considers the mass history of the subhalo. We find that our new model significantly outperforms both of the other models. We then investigate the predictive power of each input by looking at feature importance scores from the ERT algorithm. We produce feature importance plots for each baryonic property, and find that they differ significantly. We identify low redshifts as being most important for predicting star\renewcommand{\thefigure}{2a} formation rate and gas mass, with high redshifts being most important for predicting stellar mass and metallicity, and consider what this implies for nature vs nurture. We find that the physical properties of galaxies investigated in this study are all driven by nurture and not nature. The only property showing a somewhat stronger impact of nature is the present-day star formation rate of galaxies. Finally we verify that the feature importance plots are discovering physical patterns, and that the trends shown are not an artefact of the ERT algorithm.
\end{abstract}

% Select between one and six entries from the list of approved keywords, don't make up new ones

\section{Introduction}
\label{sec:introduction}

The $\Lambda$ Cold Dark Matter ($\Lambda$CDM) model is widely accepted as the standard model of cosmology \citep{lcdm_1, lcdm_2}.
In this model galaxies form and evolve within dark matter halos, virialized structures that form as a result of the gravitational collapse of the perturbations in the initial density of the universe. Large N-body simulations are run in order to model how halos form and evolve \citep{n_body_sim}. As dark matter makes up over 80\% of the mass of the universe, these dark matter only simulations are sufficient to determine the large scale structure of the universe \citep[e.g.][]{lss_1, lss_2, lss_3, lss_4}.
They also provide valuable insights into halo density profiles and substructure \citep[e.g.][]{density_profile_1, density_profile_2, density_profile_3}, although baryons can also alter the structure of dark matter halos \citep[e.g.][]{cusp_core_1, cusp_core_2}.

The ability to compare the output of N-body simulations with the observed large scale structure of the universe allows for determination of cosmological parameters and provides insight into the mechanisms of galaxy formation within halos \citep[e.g.][]{somerville_dave}.
However, large N-body simulations cannot be directly compared with observations of the universe as we only observe luminous baryonic matter.
There are several common methods used to combine baryonic physics with N-body simulations.

One common  method for accounting for baryonic physics is to run full hydrodynamical simulations. These simulations include baryonic particles or cells alongside the dark matter, and include all relevant physical processes such as gas cooling, star formation, and feedback. As these processes are below the typical resolution limits of cosmological simulations, various 'sub-grid physics' prescriptions are employed to model them. Prominent examples of cosmological simulations include Illustris \citep{illustris_1, illustris_2, illustris_3, illustris_4}, IllustrisTNG \citep{tng_1, tng_2, tng_3, tng_4, tng_5}, Simba \citep{simba}, EAGLE \citep{eagle}, HorizonAGN \citep{horizon_agn}, and FiBY \citep{fiby}. 
Despite the advent of high-performance computing units there remains a large trade-off between simulation size and resolution. Upcoming surveys such as Euclid \citep{euclid} and LSST \citep{lsst} will cover $\sim$Gp$\text{c}^3$ volumes. It is not currently possible to run full hydrodynamical simulation with high resolution at this scale.

Another approach is to take the halo catalogs resulting from an N-body simulation and 'paint on' galaxies. The simplest way to do this is via subhalo abundance matching \citep[e.g.][]{sham_1, sham_2, sham_3, sham_4}. It assumes that each halo hosts one central galaxy, each subhalo hosts one satellite galaxy, and that the highest mass halo hosts the most massive galaxy, the second highest mass halo hosts the second highest mass galaxy, and so on. The galaxy stellar masses are set such that the stellar mass function is recovered.
Another method is to use the halo occupation distribution (HOD) approach \citep[e.g.][]{hod_1, hod_2}. Here the number of galaxies within a halo are usually determined by an empirical formula which takes the halo mass as its input variable. Recent HOD models also account for secondary halo properties, such as halo concentration or environment \citep[e.g.][]{hod_3, hod_4}).
A more sophisticated technique is to use semi-analytic models (SAMs). They treat baryons using an analytic prescription that is tied to the growth of the halo from the N-body simulation \citep[e.g.][]{sam_1, sam_2, sam_3, sam_4}. SAMs tend to have a large set of tunable parameters, but the computational cost is much lower than running a full hydrodynamical simulation. Running SAMs multiple times with different sets of parameters allows for an understanding of the importance of different physical processes \citep[e.g.][]{sam_5}.

In the last decade, the field of artificial intelligence and machine learning has vastly expanded, and many of the developments are now commonly used in astrophysics. For a review see \cite{review_astro_ml}.
Machine learning algorithms are capable of learning complex nonlinear relationships between input features and target variables.
Therefore a recent approach to generating galaxy catalogs from dark matter only simulations has been to employ machine learning techniques.

The first way to utilise machine learning algorithms is to predict the number of galaxies within a friends-of-friends halo. This method is similar to the HOD method, except a machine learning model is trained to predict the number of galaxies rather than by fitting a formula. \cite{fof_pred_1} were among the first to try this approach. They used support vector machines and k-nearest-neighbour regression algorithms. This work was extended and improved in  \cite{fof_pred_conv} who used convolutional neural networks that take density fields as input to predict the number of galaxies. Recently \cite{fof_pred_sym_reg} used a combination of random forests and symbolic regression to examine the galaxy-halo connection in IllustrisTNG.

The other method of using machine learning is to learn the relationship between the baryonic properties themselves and the dark matter properties of the host halo, an approach first considered by \cite{kamdar_1, kamdar_2}. In one work they used data from the Illustris hydrodynamic simulation to train their models, and in another they trained on the Munich SAM. They used various classical machine learning algorithms and found that the extremely randomized tree (ERT) algorithm performed best. \cite{dave_1} also investigated a range of algorithms by training on the MUFASA simulation and found that the ERT was the best. They included information about the local environment around the halos as input to the models and found that this improved predictions. \cite{mssm} used the number of mergers a halo had undergone as an input feature. They applied their model to a large N-body simulation and compared the resulting galaxy catalog with one generated by SAMs applied to the same N-body simulation. They found disagreement between the machine learning approach and the SAM galaxy population, but this was to be expected as the parameters of the SAM were not tuned to the hydrodynamical simulation they trained on. \cite{shaping_gas} used similar techniques to predict the shape of gas within halos, and \cite{mocking_bh} used machine learning to predict the black hole mass of high redshift halos, but included baryonic properties as input features. An alternative machine learning technique was used in \cite{moster}. Rather than training directly on hydrodynamical simulations, they used reinforcement learning to train a neural network. Training in this way means there does not need to be a direct mapping between halos and galaxies, the network is only tasked to reproduce mass functions. This means their model can be trained on observations. They found that the halo growth rate was an important feature for making predictions. The most recent work includes \cite{dm2gal} who used convolutional neural networks that take density maps as input for prediction of stellar mass, \cite{dave_2} who use an equilibrium model as input to the machine learning models to help improve predictions, and \cite{lovell} who train their model using zoom-in simulations alongside a larger periodic box. 

In this work we use machine learning algorithms to predict the baryonic properties of dark matter subhalos. We train the model on a state-of-the-art  hydrodynamical simulation. Rather than using halo properties from redshift zero combined with summary features for the halo's history such as number of mergers or formation time, we directly use the evolutionary information of  halo properties over a wide range of redshifts. This model could be used to create galaxy catalogs from any N-body simulation that has merger trees. We show how including the full growth history of the halo significantly improves the performance of the machine learning models. We examine the features that are selected as important and show how they can be used to gain insight into galaxy formation mechanisms.

The ability of this approach to probe feature importance over time holds the key to disentangle the physical drivers of galaxy properties and to inform observational survey strategies. It allows us to examine whether correlations that have been observed between galaxies and their host halos are indeed because they are directly linked, or if the correlation is simply the result of a deeper connection at a higher redshift. As our model includes information on both a galaxies initial conditions and its evolution, it is ideal for providing insight into the "nature vs nurture" debate \cite[e.g.][]{nature_nurture_1, nature_nurture_2, nature_nurture_3}. 

The remainder of this paper is organized as follows. 
In Section \ref{sec:methods}, we explain our methodology focusing on the machine learning algorithms employed and the transformation of the input data. 
In Section \ref{sec:results}, we show how including halo properties from high redshift improves predictions and compare the performance of different algorithms. We also examine which input features our model selects as important.   
We discuss the potential of this method in section \ref{sec:discussions}, and briefly summarize the paper.

\begin{figure*}
    \centering
    \includegraphics[width=.9\textwidth]{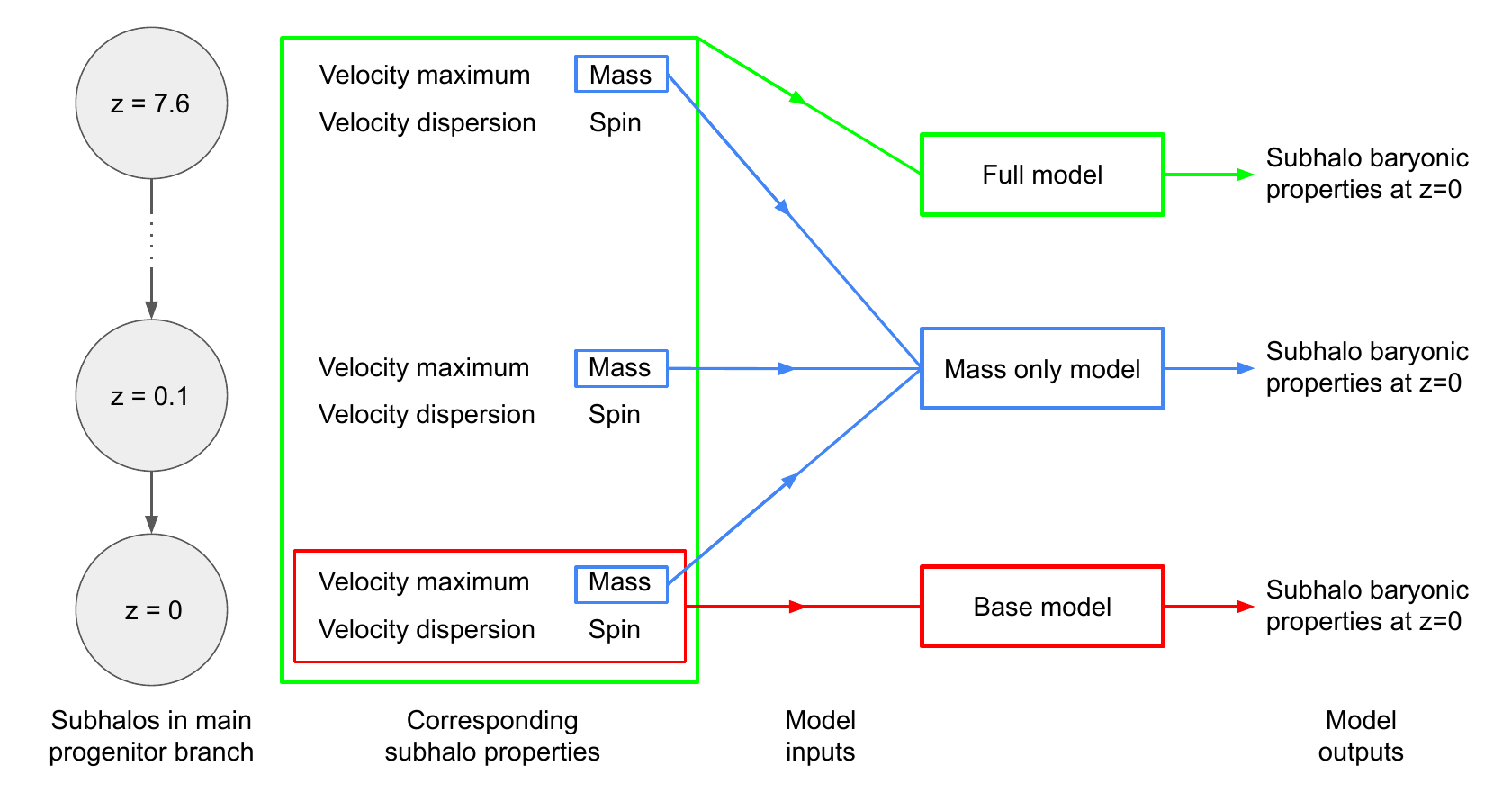}
    
    \caption{Summary of the inputs to each of the three models discussed in this work. The input features for the base model are four dark matter subhalo properties (mass, velocity dispersion, maximum circular velocity, spin) at redshift zero. The mass only model takes in the dark matter mass of the subhalo over a range of snapshots. The full model takes in the four input features of the base model, but from a range of snapshots, not just redshift zero. The output for all models is the subhalo's baryonic properties at redshift zero.} The ERT algorithm is used for all models.
    \label{fig:models}
\end{figure*}

\section{Methods}
\label{sec:methods}

In this section we provide a summary of the hydrodynamical simulation used to train our models and also an overview of the machine learning algorithms we used. We discuss how the data from the simulation must be transformed before it is possible to pass it to the models.

\subsection{Simulation}
\label{subsec:simulation}

IllustrisTNG \citep{tng_1, tng_2, tng_3, tng_4, tng_5} is a suite of hydrodynamical cosmological simulations run with the moving mesh code Arepo \citep{arepo}. Each simulation includes all significant physical processes to track the evolution of dark matter, cosmic gas, luminous stars, and supermassive blackholes from a starting redshift of $z=127$ to the present day $z=0$. 
All the simulations are run with a flat cosmology consistent with \cite{planck}: $\Omega_{\text{m},0}= 0.3089, \Omega_{\Lambda,0} =0.6911, \Omega_\text{b,0} = 0.0486, \sigma_8 = 0.8159, n_{\text{s}} = 0.9667$, and $h=$0.6774. For more details regarding the specific implementation of the IllustisTNG simulations, including all relevant subgrid models, we refer the reader to \cite{tng_implementation_1} and \cite{tng_implementation_2}.

For this work we use the TNG100 simulation which has a simulation volume of $(75 \,\,h^{-1}{\rm Mpc})^3$. The TNG100 was run from the same initial condition for three resolutions. For this work we use the highest resolution run available, named TNG100-1. This run has $1820^3$ dark matter particles with $m_{\text{DM}} = 7.5 \times 10^6 M_{\sun}$ and $1820^3$ hydrodynamic cells with $m_{gas} = 1.4 \times 10^6 M_{\sun}$ at $z=127$. Halos are found first with the friend-of-friends algorithm (FOF; \cite{fof}), then subhalos are identified using the SubFind subhalo finder \citep{subfind}. Two sets of mergers trees are available. For this work we use those generated by the LHaloTree algorithm \citep{lhalotree}. The outputs of the simulation are saved in 100 snapshots.

\subsection{Data extraction}
\label{subsec:data_extraction}

In order to ensure that the subhalos we consider  are well resolved, we require that they have their total mass (dark matter plus baryons) is above $10^9$ M$_{\odot}$ at $z=0$.  
% This cut only leaves us with $\sim$20\% of the halos from the catalog. 
This mass matches with the minimum halo mass resolved in most $\sim$Gpc N-body simulations. We disregard subhalos whose stellar or gas mass is zero, as these are not our targets of interest. This leaves us with a total of ~350,000 objects, roughly 8\% of the initial subhalo catalogue. 

In order to check the performance of our model we split the data into a train set and a test set.
For this work we assume a train volume which is 70\% of the simulation volume. The effect of varying the size of the training set is shown in Appendix \ref{sec:appendix_learning_rate}. We randomly place a subbox that has a volume of 70\% of the full volume within the full box. All halos within the box are included in our training data, all halos outside are the test data. This means that the number of halos in the training set varies depending on where the subbox is placed.
When determining the hyperparameters for the models we further split the training set into a training and validation set. This is done randomly, so the validation set does not correspond to a contiguous volume.

\subsubsection{Baseline model}
\label{subsubsec:model}
We adopt a similar baseline model to \cite{mssm}. The input features to our baseline model are the following halo properties at $z=0$: dark matter mass (the total mass of dark matter particles bound to halo, multiplied by a factor of 6/5 to account for the fact that N-body simulations do not contain any baryonic mass), velocity dispersion, maximum of spherically-averaged circular velocity, and magnitude of the spin vector. We use the ERT algorithm for our baseline model.

We do not consider any environmental properties of the halo as input features.
For the full model the total number of input features is given by $n_{\mathrm{snap}}n_{\mathrm{prop}}$, where $n_{\mathrm{snap}}$ is the total number of snapshots which we use as input for the model (we settle on $n_{\mathrm{snap}} = 10$ ), and $n_{\mathrm{prop}}$ is the number of halo properties for a single snapshot (for this work $n_{\mathrm{prop}}=4$). Therefore, increasing $n_{\mathrm{prop}}$ by including environmental properties would lead to the total number of input features in the full model being significantly larger, making the feature importance plots harder to interpret. We acknowledge that including them would improve the performance of the baseline model, but stress that their inclusion would increase the performance of the full model by a similar amount. As the purpose of this work is to demonstrate the value of taking in the full halo history, this decision is justified.

A major advantage of using decision tree based machine learning models is that they are invariant to the scaling of the input features. Therefore we do not scale the input features in any way, despite the fact that they values span multiple orders of magnitude.

\begin{figure}   
    \centering
    \includegraphics[width=.47\textwidth]{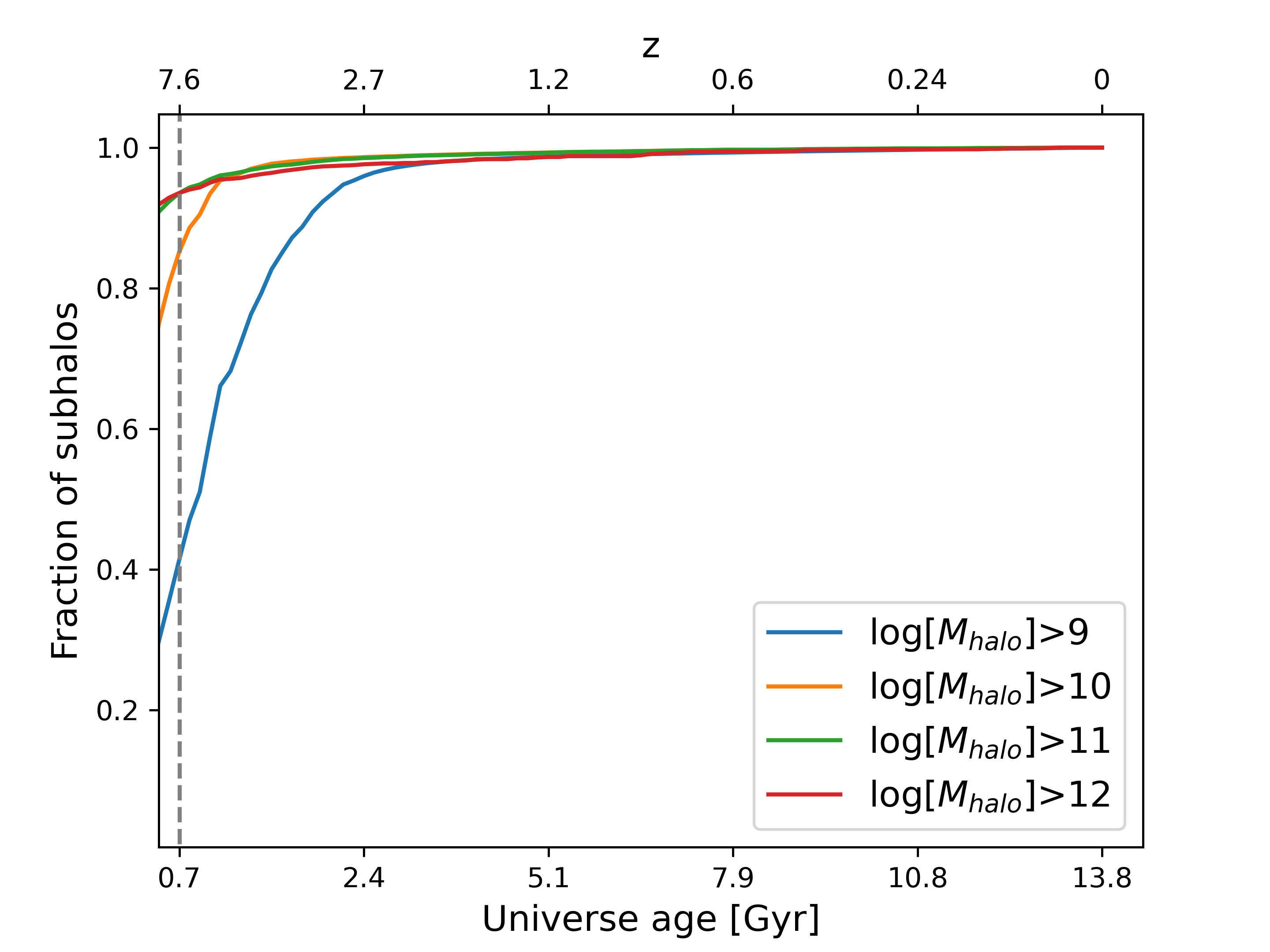}
    
    \caption{Fraction of subhalos whose merger trees extend to a higher redshift than the value on the horizontal axis. The vertical dashed line shows the highest redshift halo properties we use in this work.}
    \label{fig:fraction_subhalos}
\end{figure}

\subsubsection{Time series history}
\label{subsubsec:halo_history}

The LHaloTree algorithm constructs merger trees based on subhalos rather than FOF halos. For a given subhalo in a given snapshot, the algorithm finds all subhalos in the next snapshot that contain some of its particles. These are counted in a weighted fashion, with a high weight assigned to more tightly bound particles. The subhalo with the highest count is selected as the descendent.
% Minimum number of particles

For each valid subhalo at $z=0$ we track its properties back in time using its main progenitor at each snapshot. At each snapshot to be used as input for the model we store the same four properties that we use for our baseline model. All these features are passed as input features to our full model which predicts the subhalo's $z=0$ baryonic properties. We also have a model which only takes in the subhalo mass at the different snapshots being considered. We refer to this as the mass only model. Figure \ref{fig:models} shows a summary of each of the three models.

To allow for subhalos that temporarily disappear, the LHaloTree algorithm may link a subhalo identified at snapshot $n$ with one at snapshot $n-2$ if no progenitor can be found at snapshot $n-1$. 
Therefore a subhalo may be missing properties at a point in its merger tree. Whenever a subhalo is identified at snapshot $n-2$ and snapshot $n$, we set the subhalo properties at snapshot $n-1$ as equal to the values at snapshot $n$.

The TNG100 simulation has halo properties stored for 100 snapshots, with snapshot 99 corresponding to $z=0$.
Figure \ref{fig:fraction_subhalos} shows the fraction of subhalos that can be tracked to at least the snapshot shown. Higher mass halos are easier to track further back.
% The highest mass can't always be tracked because they come from such a dense environment
We therefore decide the lowest snapshot to consider is 9, which corresponds to $z=7.6$. 
If a halo cannot be tracked back to a certain redshift the value of its input features for that snapshot are set to zero. In that case the performance of the model is worse than for the halos we can track back, but it is still better than our baseline model. As can be seen from Figure \ref{fig:fraction_subhalos}, a significant fraction of subhalos can be tracked back to $z=7.6$, so there is still a benefit to using the halo properties at this redshift as input to our models.

\subsubsection{Output features}
\label{subsubsec:output_features}

Although the model can be used to predict any baryonic property of a subhalo, in this work we focus on 8 properties: the gas mass, the total black hole mass, the stellar mass, the mass weighted metallicity of the star particles, the stellar half mass radius, the sum of the star formation rate of all gas cells, and the U and K band magnitudes. Following \citep{mssm}, we log all values except for the magnitudes. If we do not take the logarithm, during the training phase the data points from high mass halos are weighted much more highly than those from low mass halos. We scale each output feature using a MinMax scaler (subtract minimum value, divide by maximum - minimum value) to transform all values to be between 0 and 1. We use the mean squared error to evaluate the performance of our models.

\begin{equation}
    \text{MSE} = \frac{1}{n} \sum_{i=1}^{n} ( y_i - \hat{y_i} ) ^ 2
    \label{eq:mse}
\end{equation}

Here $n$ is the number of data points, $y_i$ is the true value of the output feature, and $\hat{y_i}$ is the value predicted by our models. Due to the normalization of the output features the value of the mean squared error does not have a physical significance. However this rescaling allows us to compare how difficult each output feature is to predict.

\subsection{Machine Learning Methods}
\label{subsec:ml_methods}

We exploit the results of fully hydrodynamic, high-resolution simulations to create a mapping between halo and galaxy properties. This type of problem is known as supervised learning as we have a set of input data (dark matter only properties of a subhalo) and corresponding output data (baryonic properties of the galaxy hosted by that subhalo). Below we give an overview of the supervised machine learning algorithm that we use.

\subsubsection{Extremely randomised tree ensembles}
\label{subsubsec:rf}

Random forest regressors \citep{rf_1,rf_2} use an ensemble of decision trees to make a prediction. Each decision tree is constructed top-down from a root node. At each node the data is split into two bins based on the values of its input parameters. The splits are chosen such that the weighted average of the MSE of the two bins is minimized \citep{gini}. This partitioning of the data results in each leaf node at the bottom of the tree containing a small subset of the data, where almost all members of the subset have a similar output value. Predictions from decision trees are based on the assumption that test data points will have a similar output value to the other members of the leaf node it is placed into. A random forest is made up of a number of decision trees. There is a bootstrapping procedure such that each decision tree within the forest is trained on a randomly generated subset of the training data. Further randomness is added in that for each split only a subset of input features can be used. The prediction from a random forest is the average prediction of its component decision trees. A major advantage of random forests is that they are significantly less prone to overfitting data compared with a single decision tree. This results from the randomness added when training the individual decision trees.

For this work we use extremely randomised tree ensembles (ERT; \cite{ert}). This is the algorithm used in previous work \citep{kamdar_2, dave_1, mssm}, and we found it to slightly outperform the standard random forest. It adds in additional randomization by computing a random split for each feature at each node, rather than the optimal split.

\subsubsection{Determining hyperparameters}
\label{subsubsec:hyperparameters}

Machine learning algorithms often have hyperparameters. These are parameters of the model itself, and the values do not change when the model is trained. They control properties of the model such as its complexity, or how fast it learns. For the ERT we consider different values for the \textit{n\_estimators}, \textit{max\_depth}, \textit{min\_samples\_leaf}, and \textit{min\_samples\_split}. We retain the default values for the other hyperparameters.
The value of \textit{n\_estimators} sets the number of decision trees within the ERT. With too few decision trees the model will have a tendency to overfit the data. Having too many trees should not decrease the performance of the model, but it will increase the time the model takes to run and make predictions.
The value of the \textit{max\_depth} hyperparameter limits how many nodes there can be in each decision tree. This constrains the maximum number of input features each decision tree can use, since each depth splits on only one input feature. If the value of \textit{max\_depth} is too high the model may be prone to overfitting.  
The \textit{min\_samples\_leaf} and \textit{min\_samples\_split} hyperparameters combine to specify the minimum number of data points a node must contain in order to split into further nodes. By increasing the value of these parameters, we can decrease the total number of splits. This limiting of the number of parameters in the model can further prevent overfitting. The hyperparameter values found are shown in Appendix \ref{sec:appendix_bayes_opt}. 

Picking the values for the hyperparameters can be seen as a black box optimization problem, where the objective function to be minimized is the performance of the model on a test data set. Common methods for tuning hyperparameters include random search and grid search. For this work we use Bayesian optimization. After evaluating the performance of the model for a small set of randomly chosen hyperparameters, a prior distribution is calculated to capture beliefs about the behaviour of the objective function. From this an acquisition function is calculated that determines the next values of hyperparameters to try and evaluate. For more details see \cite{bayes_opt}. More information about our implementation is given in Appendix \ref{sec:appendix_bayes_opt}.

\subsubsection{Feature importance}
\label{subsubsec:method_feature_importance}

One major benefit to using ensembles based on decision trees is the ability to extract information on which input features are providing the information that is used to make the final predictions. For each decision tree the importance of an input feature can be determined by the number of times it is used for a split, and how close to the top of the tree those splits are. By averaging the importance values over all the decision trees within the ensemble model, a set of the feature importances can be determined for the model as a whole. However, one must be aware that correlations between input features will affect their importance values, and can make the results more difficult to interpret. The sum of the feature importance of all input features is normalised to one. Therefore when examining feature importance plots the differences in the relative importance of each input feature should be considered, rather than their absolute values.

After training a model to predict a single baryonic property, we will look at the feature importance to establish which input features contribute most to determining the value of the output feature. As our input features span a range of redshifts, peaks in the feature importance will tell us which times in a galaxies evolution are most important for setting the final value of each baryonic property. It should be noted that a high feature importance value does not tell us if an input feature is positively or negatively correlated with the output feature.

\section{Results}
\label{sec:results}

\begin{figure}   
    \centering
    \includegraphics[width=.47\textwidth]{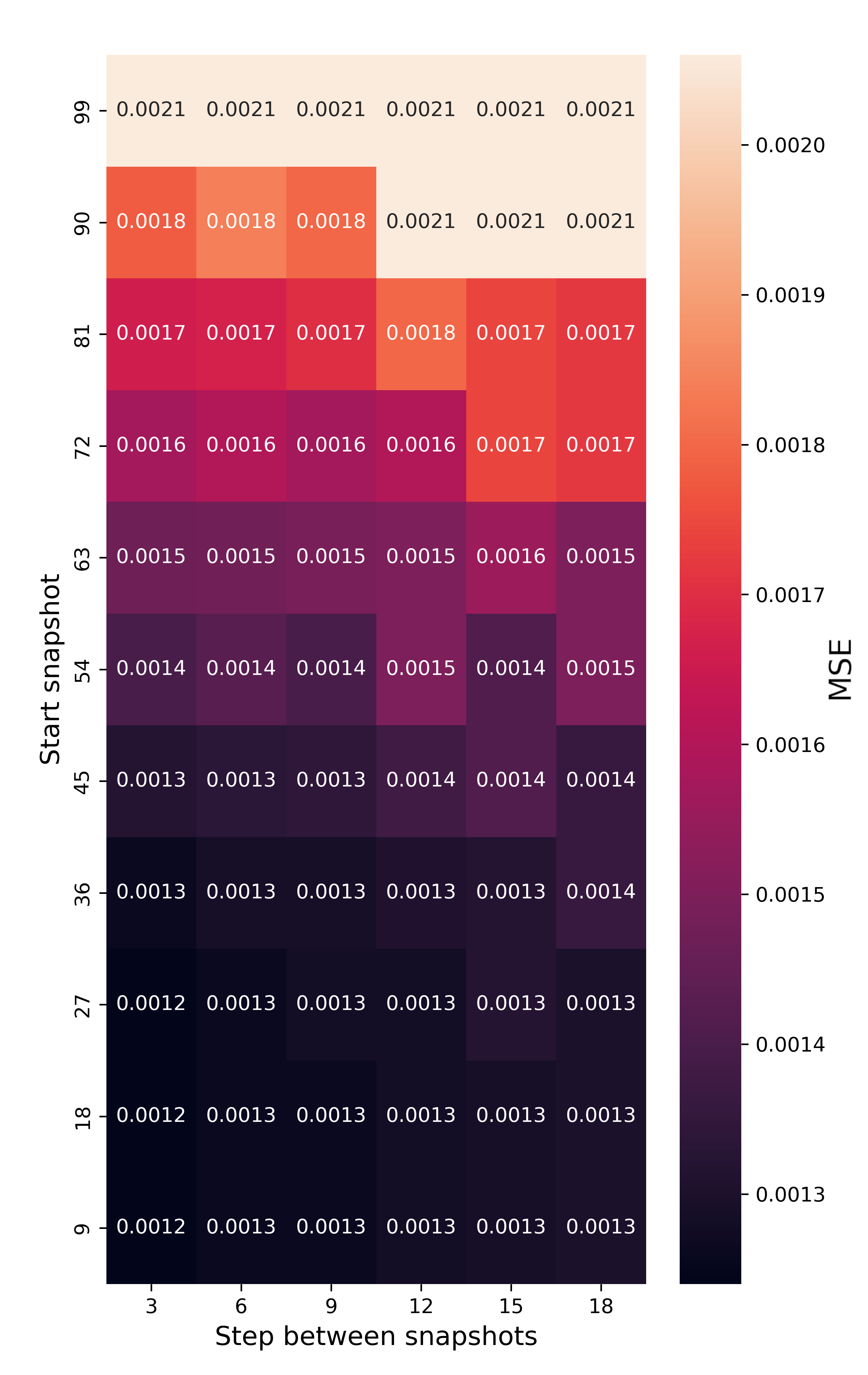}
    
    \caption{Performance of the regressor for different snapshot ranges. A lower MSE score indicates more accurate predictions. The start snapshot values indicates the highest redshift halo properties passed to the machine learning model. The model performs more accurately as the start snapshot decreases, showing how including halo properties at high redshifts is beneficial.}
    \label{fig:snapshot_mse_matrix}
\end{figure}

\begin{table*}
    \centering
    \caption{The mean squared error, Eq. \ref{eq:mse}, quantifying the performance of different models at predicting baryonic properties of subhalos. All scores are for predictions on the test set, aside from the final row.}
    \begin{tabular}{|c|c|c|c|c|c|c|c|c|c|}
                                     & BH mass & Gas mass & Half mass radius & U band & K band & SFR    & Stellar Mass & Stellar Metallicity 
        \\ \hline \hline
        Mass only model              & 0.0017  & 0.0019   & 0.0027           & 0.0024 & 0.0018 & 0.0064 & 0.0015       & 0.0073 
        \\ \hline
        Baseline model               & 0.0019  & 0.0021   & 0.0028           & 0.0029 & 0.0026 & 0.0061 & 0.0021       & 0.0081 
        \\ \hline
        Full model                   & 0.0012  & 0.0017   & 0.0025           & 0.0019 & 0.0016 & 0.0049 & 0.0012       & 0.0069  
        \\ \hline
        Full model (Train)           & 0.0010  & 0.0017   & 0.0024           & 0.0018 & 0.0015 & 0.0045 & 0.0012       & 0.0068  
        \\ \hline
    \end{tabular}
    
    \label{table:model_performance}
\end{table*}

\begin{figure*}   
    \centering
    \includegraphics[width=\textwidth]{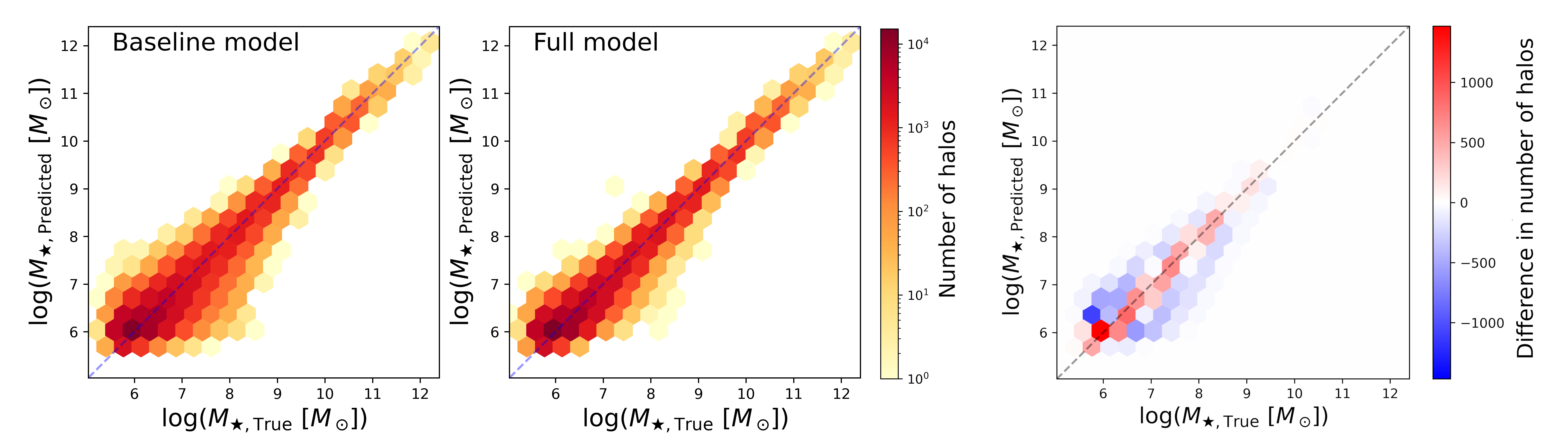}
    
    \caption{(Left) A hexbin plot showing stellar mass values predicted by the baseline model compared with their true values. The blue dashed lines corresponds to a perfect prediction. (Middle) Same as left plot, except predictions are from the ERT model that takes in the halo properties from 10 snapshots, starting at redshift $z=7.6$. Both plots are generated from the same train/test split. The scatter is reduced compared with the left plot, indicating an improvement in prediction accuracy. (Right) Difference in number of halos in each bin between the left and middle panels. Positive value indicate that the improved model has more halos in that bin than the baseline model.}
    \label{fig:stellar_mass_hexbin}
\end{figure*}

\subsection{Which snapshots to include}
\label{subsec:which_snapshots}

We use the ERT model and vary the snapshots that we include to see what effect it has on the performance of our model. The results when predicting the stellar mass of subhalos are shown in Figure \ref{fig:snapshot_mse_matrix}, where a lower MSE score indicates better performance. The start snapshot represents the highest redshift snapshot passed as input to the model, and the step between snapshots gives the spacing. For example a start snapshot of 72 with a step between snapshots of 9 would correspond to the halo properties at snapshots 72, 81, 90, 99 being fed as inputs features. The values for the MSE are the average of 10 different training/test splits. As the maximum snapshot from IllustrisTNG is 99, the top row corresponds to the baseline model, where only the $z=0$ properties are passed as input. It can been seen that starting at a lower snapshot improves the performance of the model. This improvement continues down to our lowest starting snapshot of 9, which corresponds to $z=7.6$. Decreasing the jump between snapshots also causes the MSE to decrease, but the effect is much smaller than varying the starting snapshot. Setting the step size to 1 (not shown here) does not improve the model compared with a step size of 3. Similar figures showing the same trends are produced when predicting baryonic properties other than the stellar mass. From this figure we choose a starting snapshot of 9 with a step size of 10 as the full model for the rest of this work. We wish to start with the lowest possible snapshot to gain the most predictive power. We note that there is no disadvantage to starting with the lowest snapshot, even though not all subhalos can be tracked back to this point. We chose the larger step size as it is easier to interpret the feature importance plots when there are fewer input features, as discussed in section \ref{subsubsec:model}.

% Do we need to reach the SFR peak?
% For TNG the peak is at z=3, see https://arxiv.org/abs/1703.02970

\subsection{Comparison of models}
\label{subsec:model_comparison}

The results of our different models are shown in Table \ref{table:model_performance}. A lower MSE score indicates better model performance. The values given are the average of 10 different training/test splits. The standard error on the mean from the 10 runs is of the order $10^{-5}$ and so is not shown.  We can see that in all cases the full ERT model outperforms the baseline model. We indicate this visually in Figure \ref{fig:stellar_mass_hexbin}. In the left and middle panel we plot the stellar mass values predicted by the baseline and full models respectively, compared with their true values. It is clear to see that the scatter in values has decreased for the full model. To highlight this in the right panel we plot the residual of the left and middle panels. This shows that not only is there reduced scatter, but the full model has a larger number of predictions lying on the diagonal, indicating a correct prediction.

Since we have normalized the output features it is possible to get an idea of how difficult each feature is to predict by directly comparing MSE scores. We see that SFR and stellar metallicity are the most difficult to predict, and this is in agreement with previous work \citep[e.g.][]{kamdar_2}. 
% However their conclusion came from looking at the scatter when plotting the values rather than by directly comparing MSE values.
The reason that SFR is difficult to predict is due to it's stochasticity.
Stellar mass, gas mass, and black hole mass are all integrated quantities which build up over time as gas falls onto the subhalo and is processed. This explains why they are easier to predict, as the stochastic processes involved in their rate of change are smoothed out by considering a large range of snapshots.
The stellar metallicity is dependent on a number of complex factors, such as when the bulk of star formation took place, how much recycling of metals produced by previous star formation there was, and how much unpolluted fuel is available to the galaxy. The interplay of these numerous processes explains why the MSE is higher for stellar metallicity predictions than for any other output feature.
% The stellar metallicity is dependent on when the star formation that formed the bulk of the stars took place. If the majority of star formation takes place early the metallicity will the low, but if it takes place at late times this will give a high metallicity. As SFR is difficult to predict, this means the metallicity MSE is large compared with the other output properties.
% \adg{(SK: not sure I agree. E.g. massive early type galaxies form most of their stars early and have also large metallicities. Metallicities correlate well with the mas sof a galaxy, the so-called mass metallicity relation. Thus it is not just when star formation took place but the ammount of recycling of metals vs the provision of unpolluted fuel etc. It is more complex. )} 
The U band and K band magnitudes are strongly linked to the number of stars giving out light, i.e. the stellar mass. The MSE score for the U band is higher as it is linked to young stars and falls off quickly over time, and so it is more closely associated with the current SFR of the galaxy than the K band.
% Not sure what to say about galaxy size
% Although the MSE scores are very different for metallicity and stellar mass, their feature importance plots are very similar. This suggests that could be another output feature which we are missing which would be very important for metallicity predictions, and less important (but probably still helpful) for stellar mass predictions.

The improvement in score between the baseline model and the full model gives an indication of how important a subhalo's history is in determining the value of a certain property. We expect there to be a larger improvement for properties that are more dependent on the exact growth and merger history of the halo. For example the baseline model gives the same MSE score for both stellar mass and gas mass. Using the full model gives a much greater improvement to the stellar mass prediction than the gas mass prediction. This is to be expected as for most subhalos a significant fraction of their stellar mass was created at high redshifts, whereas the gas found in a subhalo at early times may be used up in star formation or blown out by feedback. The MSE is always lower for the full model than for the mass only model. This shows how it is important to include other properties of the subhalo at higher redshifts than just its mass history. When comparing the results of the mass only model with the baseline model we see that for most baryonic property predictions the mass only model is better. However for predictions of the SFR the baseline model is better. This is because SFR is the only property we predict that is instantaneous, rather than being built up over time (e.g. stellar mass). In this case the machine learning mode finds it preferable to have as much information as possible about the $z=0$ subhalo properties when predicting an instantaneous property. 
These results give the first indication that our models have shown nurture to be more important than nature. The full model can be linked to nurture, as it takes into account the evolution of the halo. The baseline model can be linked to nature, as it takes information about the halo at a single point in time. If nature was more important than nurture in determining a halo's properties, the model would be able to approximate the link between the halo properties at $z=0$, and halo properties at the snapshot that defines the halo properties. However, as we see the full model always significantly outperforms the baseline model, then this cannot be the case.

In the final row of Table \ref{table:model_performance} we show the MSE of the full model on the training set. For most for the output features the MSE of the test set is slightly larger than the training set, but the difference is small enough that it shows our model is not overfitting.

% Standardised MSE can be used in conjunction with subsampling

\begin{figure*}
    \centering
    \includegraphics[width=.38\textwidth]{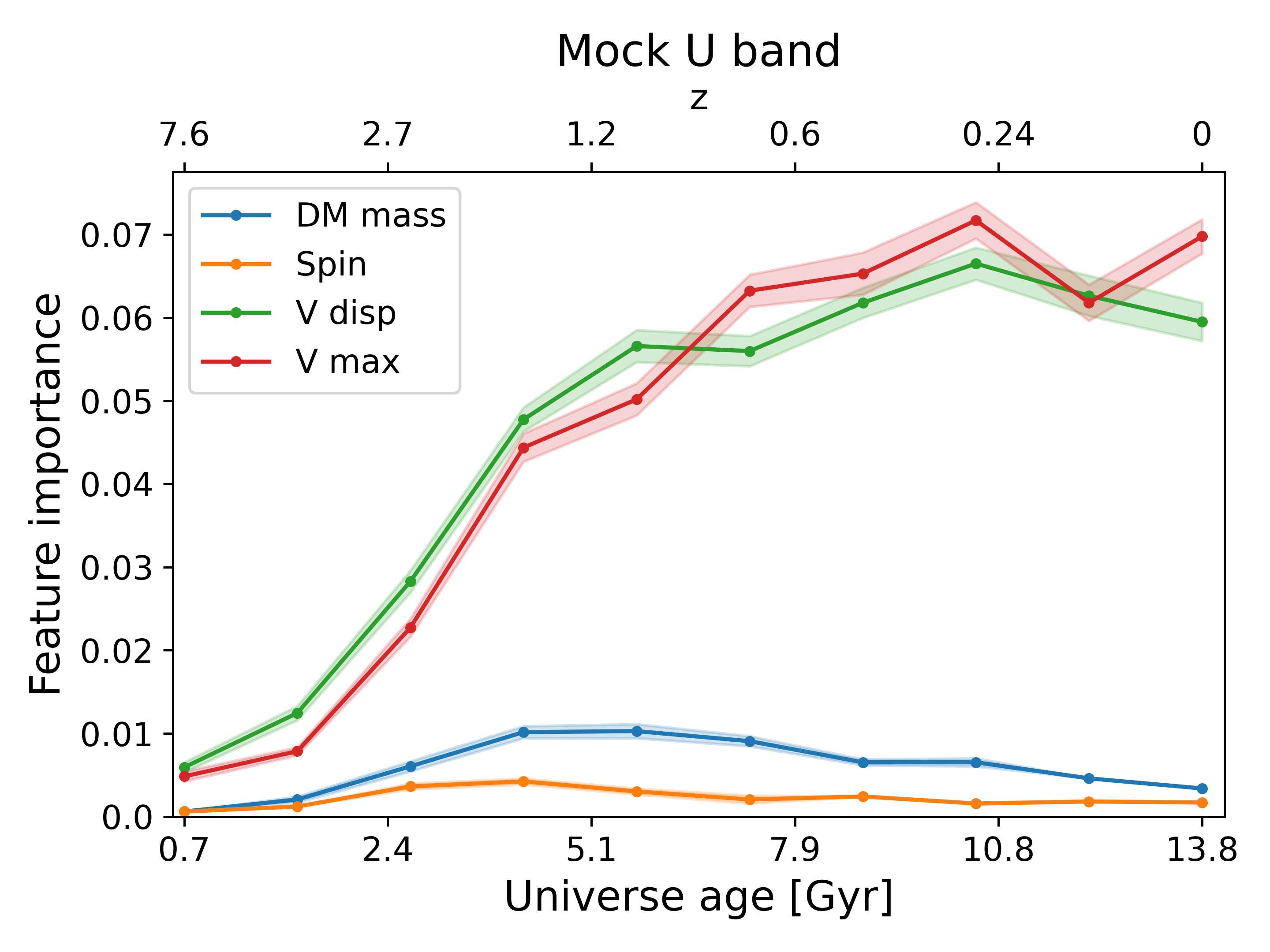}
    \hspace{1cm}
    \includegraphics[width=.38\textwidth]{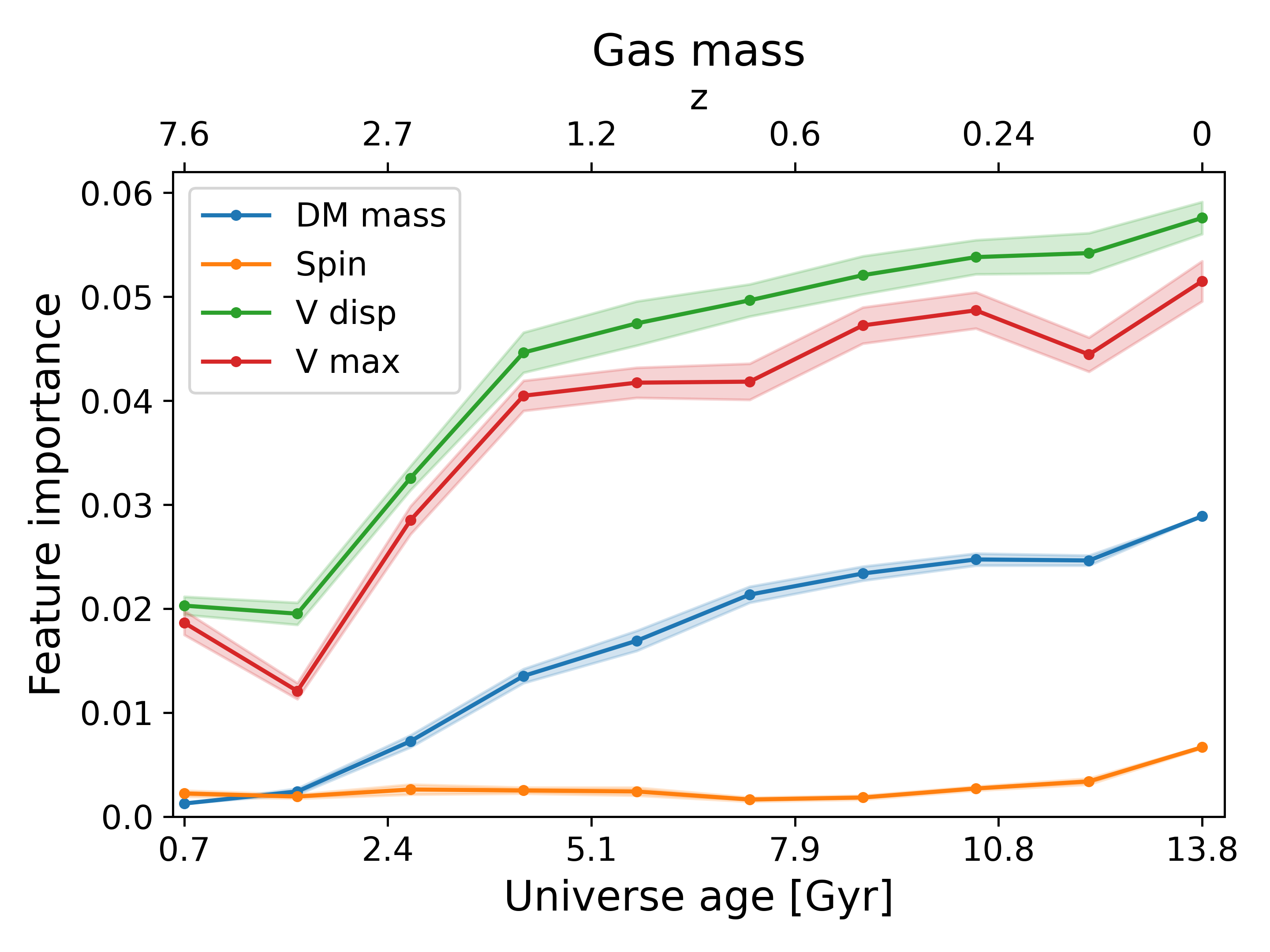}
    
    \vspace{5mm}
    
    \includegraphics[width=.38\textwidth]{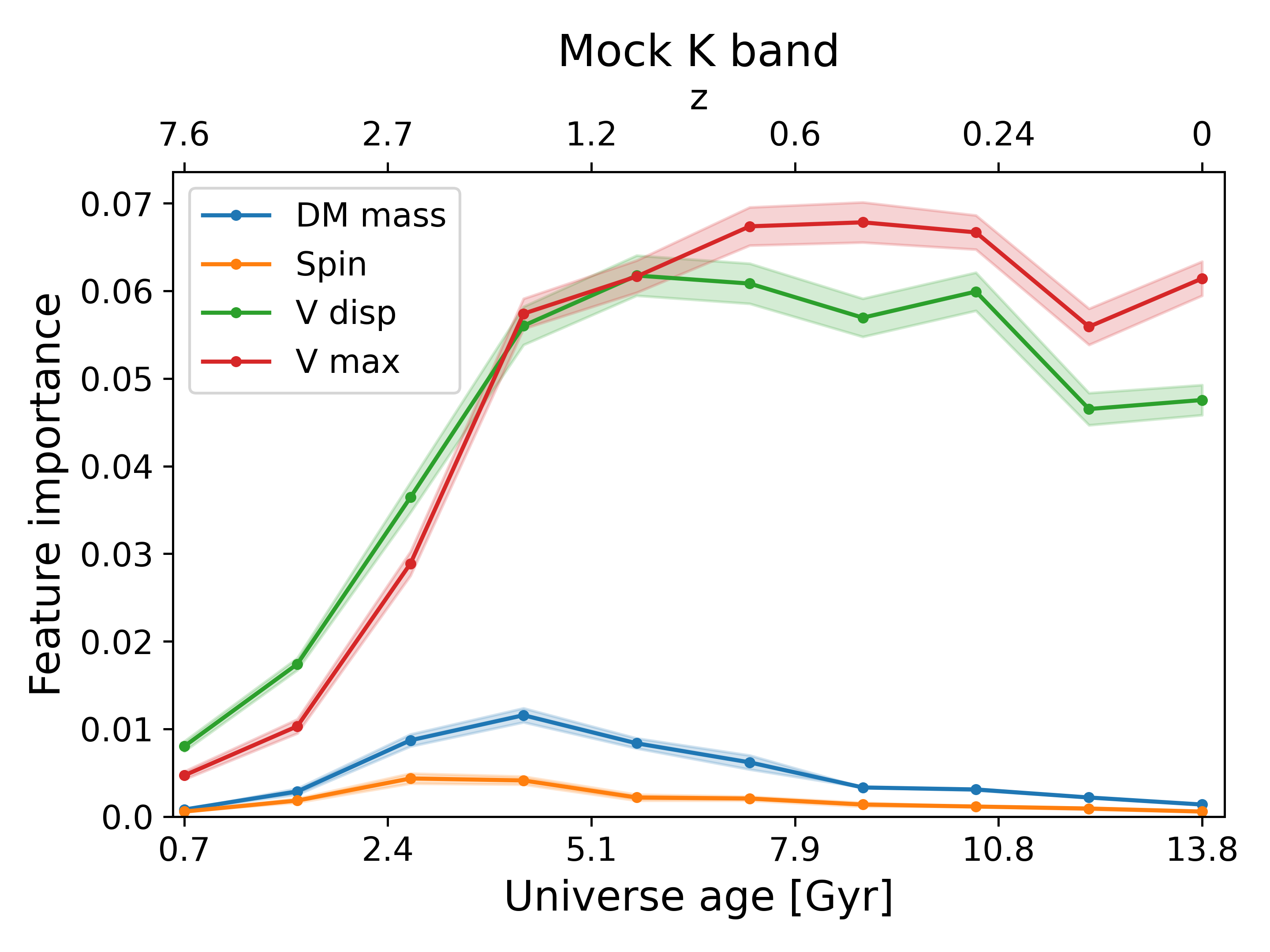}
    \hspace{1cm}
    \includegraphics[width=.38\textwidth]{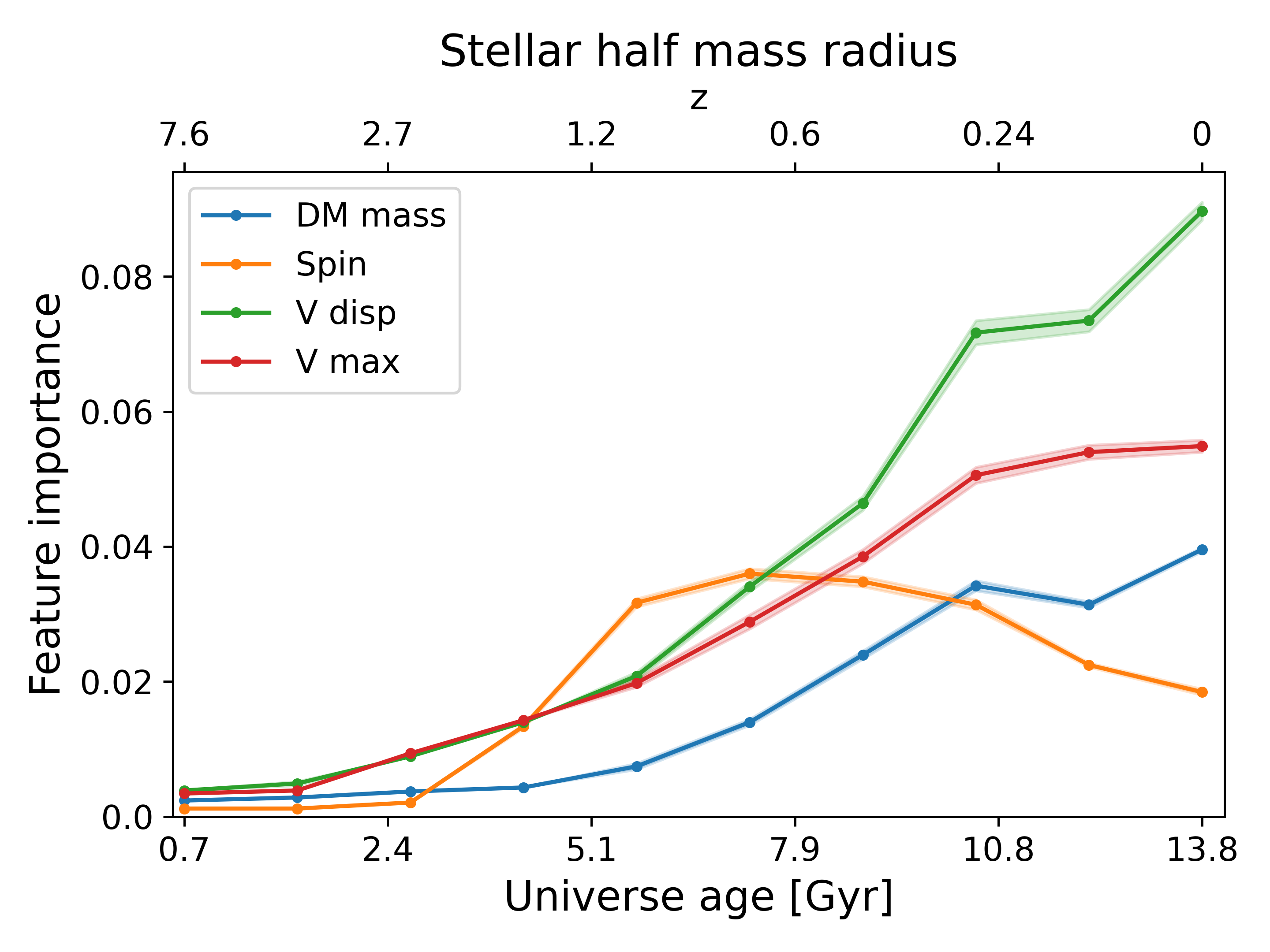}
    
    \vspace{5mm}
    
    \includegraphics[width=.38\textwidth]{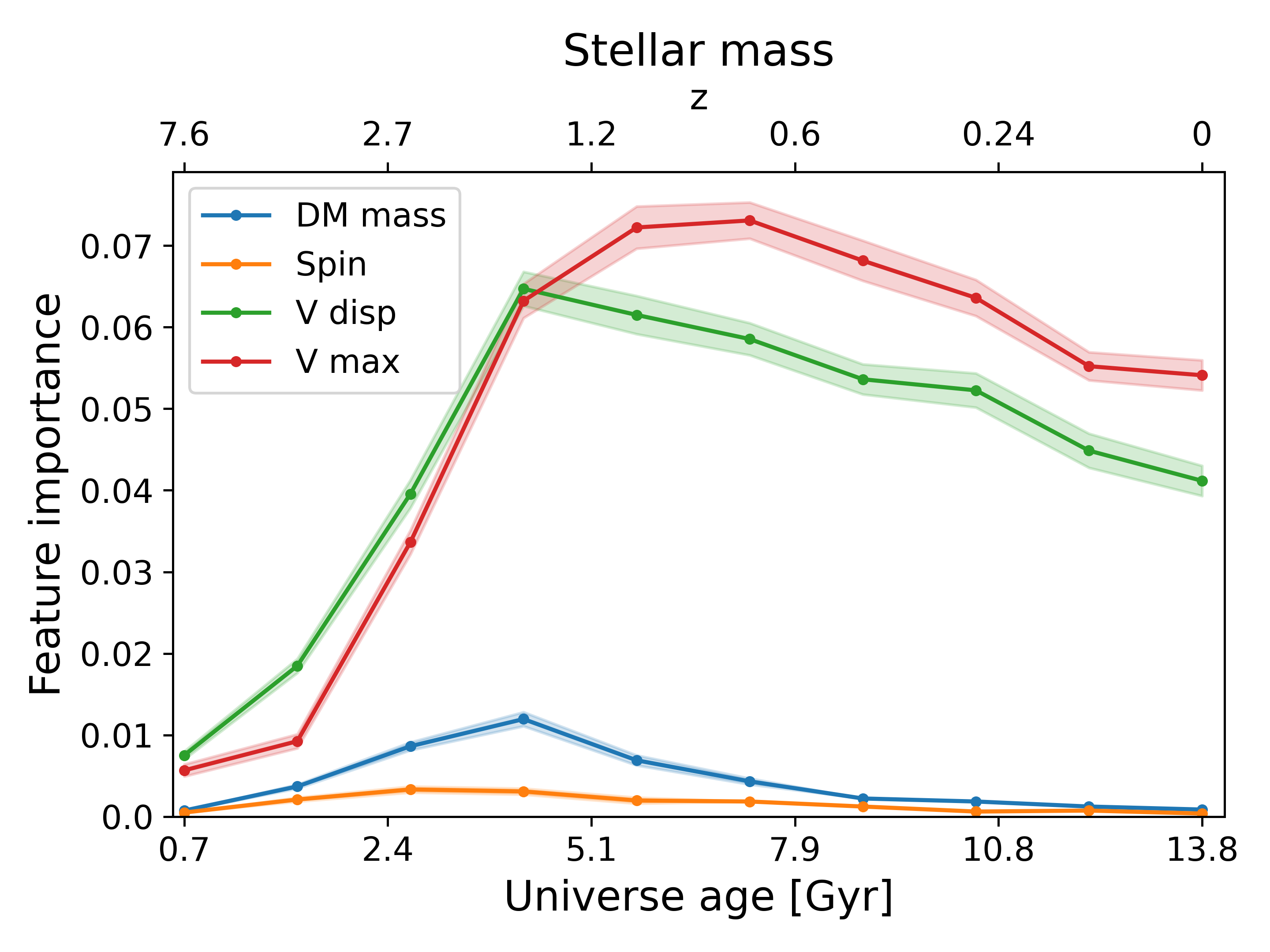}
    \hspace{1cm}
    \includegraphics[width=.38\textwidth]{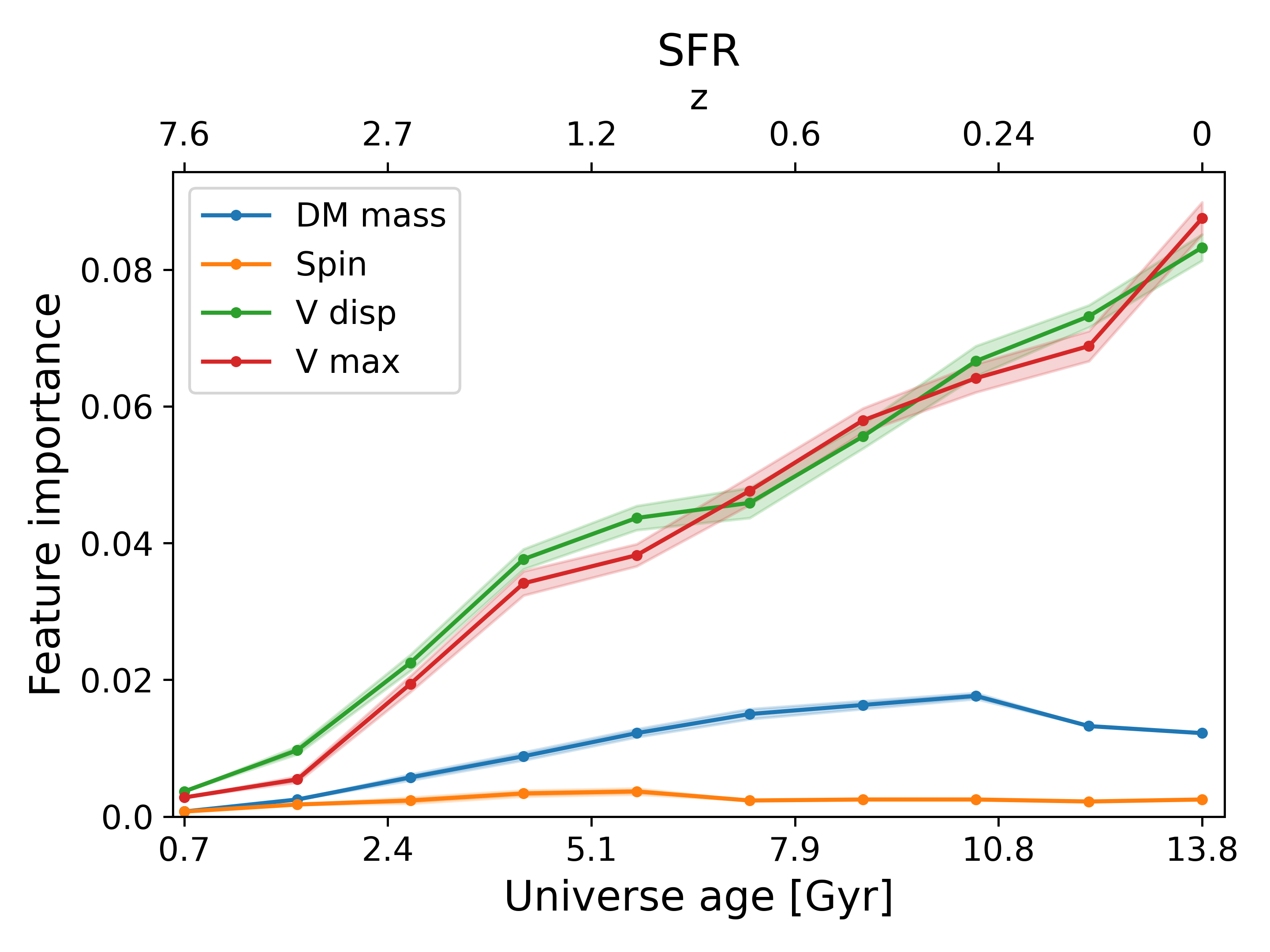}
    
    \vspace{5mm}
    
    \includegraphics[width=.38\textwidth]{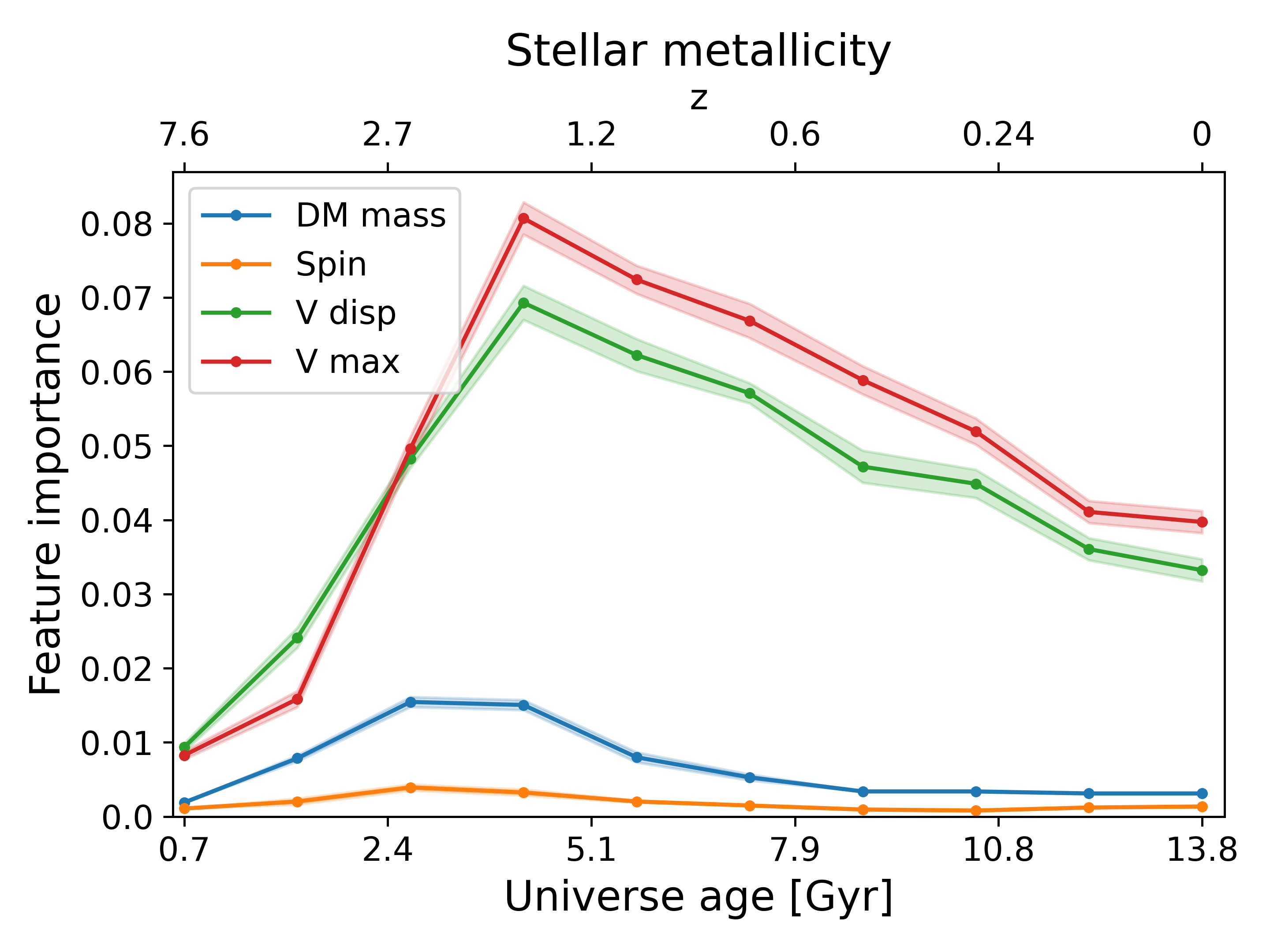}
    \hspace{1cm}
    \includegraphics[width=.38\textwidth]{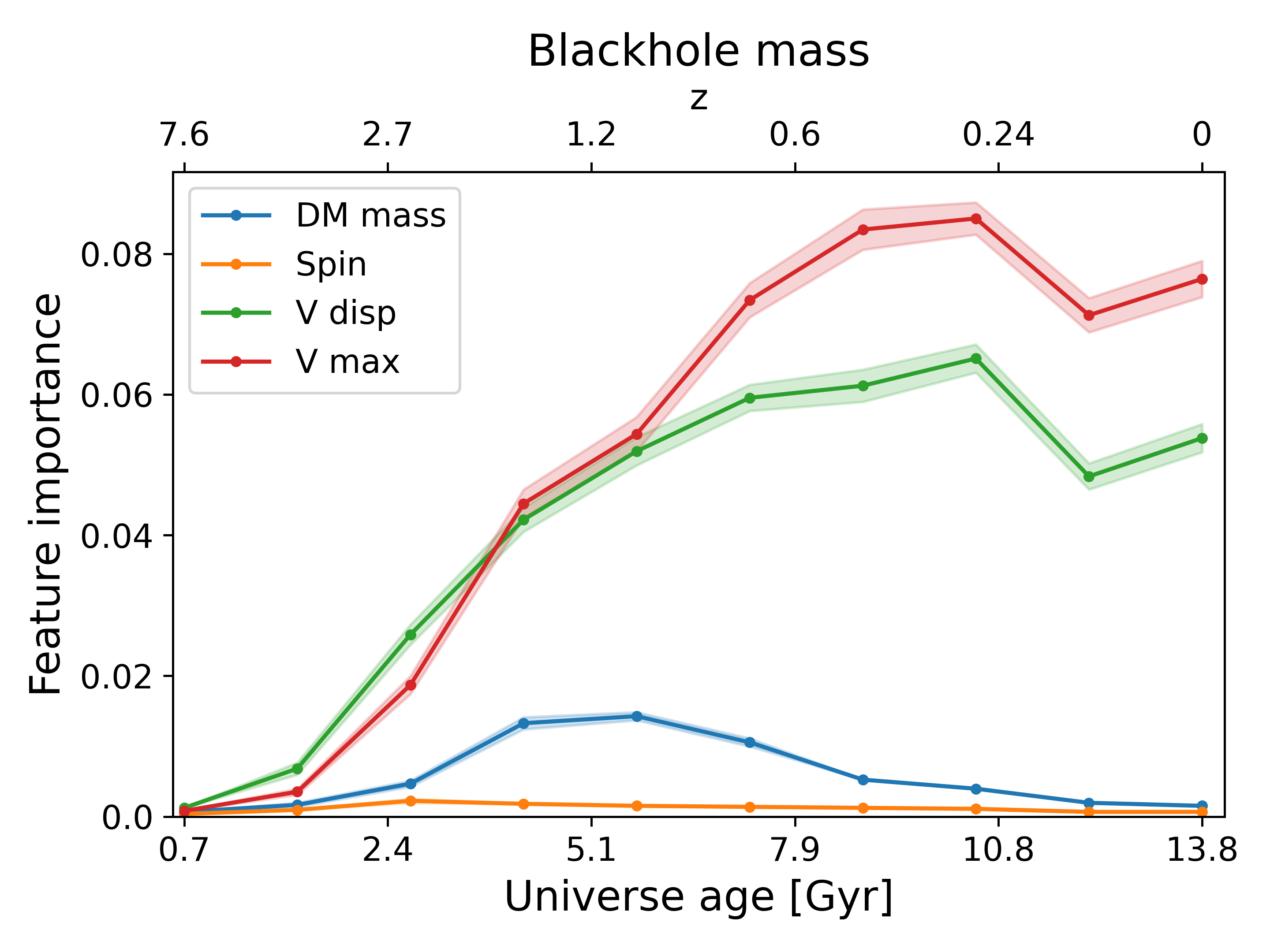}
    
    \caption{Feature importance values from the ERT model that takes in the halo properties from 10 snapshots, starting at redshift $z=7.6$. The shaded region represents one standard error in the mean  based on training and evaluating the model 10 times. The variation results from a combination of the different training sets used each time, and from the inherent randomness in the ERT algorithm.}
    \label{fig:feature_importance}
\end{figure*}

\subsection{Feature importance from ERT models}
\label{subsec:feature_importance}
In Figure \ref{fig:feature_importance} we show the plots of the feature importance values of each input feature fed into the full model. As we train a separate model for each baryonic property being predicted we end up with 8 different plots of feature importance. The feature importance is calculated as described in section \ref{subsec:ml_methods}. The sum of the feature importance over all features in a model is normalised to be equal to one. Each point on the plot represents one input feature to the full model. The shaded region is the $1 \sigma$ standard error in the mean from 10 different training/test splits. A test set is not needed to calculate the feature importance, but we train with 10 different splits to get an estimate of the variation. It is clear from the contrasting plots in Figure \ref{fig:feature_importance} that the feature importance varies significantly depending on the output feature which is being predicted. This shows our models are picking up on the different ways each baryonic property is built up. There are some common trends, mainly that the halo velocity dispersion and halo maximum velocity are determined to be the most important features. 
Both \cite{kamdar_1} and \cite{dave_1} looked at feature importance, but used similar models to our baseline model, i.e. they did not consider the full halo growth history. They also did not train a separate model for each output feature, so the feature importance values found were not associated with a single $z=0$ galaxy property. In agreement with our overall trends, they also found velocity dispersion and maximum velocity to be good predictors. Also in concurrence with previous work we find that in general spin is the feature which provides the least information. The fact that for all plots there are times where the spin feature importance goes to zero is evidence that our models are not over-fitting. 

The first major difference between the feature importance plots is the location of the peak. Comparing the peak of the feature importance plots gives an indication of what epoch is most important for the build up of that baryonic property. Unsurprisingly SFR peaks at $z=0$. This agrees with the discussion in section \ref{subsec:model_comparison} about the score difference between the mass only and baseline model. Although feature importance plots such as Figure \ref{fig:feature_importance} can only be obtained from decision tree-based machine learning models, similar MSE scores to the ones shown in Table \ref{table:model_performance} are obtained if we use a different algorithm. The fact that our feature importance plots agree with the model-agnostic MSE results is a confirmation that the feature importance are able to determine physically interesting results.

The feature importance plots for stellar mass and stellar metallicity are similar. This is to be expected as metallicity of the stellar particles is strongly correlated with the time at which the stellar mass is formed. For the IllustrisTNG simulations the peak in cosmic star formation rate density occurs around $z=2$. We might expect the stellar mass feature importance to peak at the same point, but it appears later, around $z=1$. 
% Reasons for this. Units of SFRD?
The stellar mass and K band magnitude plots are similar. K band magnitude is often used as a proxy for stellar mass \citep[e.g.][]{k_band_1, k_band_2, k_band_3, k_band_4}, and our models are able to independently pick up on this connection. Comparing the feature importance of the K and U bands shows a peak at different times. The U band peaks at $z=0$ which makes sense as UV light is emitted by young stars and so is correlated with SFR.
% Peak in dm mass feature importance

Looking at the plots for SFR and for gas mass, we see that in both cases halo velocity dispersion and halo maximum velocity are most important but the importance of dark matter mass differs. For SFR the peak in dark matter mass is prior to $z=0$. As the dark matter mass determines the gravitational potential of the subhalo it will be linked with the amount of infalling gas. However this gas can only be used for star formation once it has cooled, whereas the gas mass of the subhalo includes both hot and cold gas. This explains why the peak in dark matter mass feature importance does not occur at $z=0$ for SFR, but does for gas mass.

\begin{table*}
    \centering
    \caption{The ratio of $I_{\rm{nat}}$ to $I_{\rm{nur}}$ for each of the output properties (top) being predicted based on input properties (first column). Values larger than 0.5 suggest nature is more important for a given physical property of galaxies, while values smaller than 0.5 support nurture as the main driver. As can be seen in the table, all values for the galaxy properties listed here are below 0.5 and thus nurture is the dominant driver of galaxy properties.}
    \begin{tabular}{|c|c|c|c|c|c|c|c|c|c|}
        F(t)      & BH mass & Gas mass & Half mass radius & U band & K band & SFR   & Stellar Mass & Stellar Metallicity 
        \\ \hline \hline
        DM mass   & 0.33    & 0.23     & 0.30             & 0.21   &  0.31  & 0.24  & 0.37         & 0.32
        \\ \hline
        Spin      & 0.21    & 0.34     & 0.23             & 0.21   &  0.32  & 0.21  & 0.22         & 0.32
        \\ \hline
        Vel disp  & 0.16    & 0.16     & 0.32             & 0.19   &  0.17  & 0.25  & 0.17         & 0.18
        \\ \hline
        Vel max   & 0.20    & 0.17     & 0.26             & 0.19   &  0.17  & 0.25  & 0.16         & 0.19
        \\ \hline
        All       & 0.18    & 0.17     & 0.22             & 0.18   &  0.16  & 0.23  & 0.17         & 0.18
        \\ \hline
    \end{tabular}
    
    \label{table:feature_importance_peak}
\end{table*}

Models of galaxy formation and evolution often assume a relation between the spin of a galaxy and its host halo \citep[e.g.][]{half_mass_1, half_mass_2}. This relation is used to set the size of galaxies in many semi-analytic models. However, recent work has suggested that this relation may not hold in cosmological simulations \citep[e.g.][]{half_mass_3, half_mass_4}. From our feature importance plot for stellar half mass radius we see that within IllustrisTNG the spin of a halo is a predictor of galaxy size, in agreement with \cite{half_mass_5}. We find that the importance of the halo spin in determining the $z=0$ galaxy size peaks around $z=1$. This shows that the galaxy size at $z=0$ is less correlated with the halo spin at $z=0$ than the halo spin at $z=1$. This suggests that at earlier times the halo spin was important in determining the angular momentum of the galaxy, and therefore the galaxy size. However, as the halo has continued to grow and evolve after the galaxy has formed the spin of the halo no longer effects the galaxy.
% This suggests that as the halo continues to grow and evolve after the galaxy has formed that their spin has decoupled.
This information could only be found because our method takes in such a wide range of redshifts. When considering the full range of redshifts we find that the other halo properties are more important than spin.
% This is becuase the size of the galaxy is also determined strongly by it's stellar mass, for which v_disp and v_max are the most important. But then why does it not peak at the same point the feature importance for stellar mass peaks?
% This could be due to the fact that there is only a strong connection between spin and galaxy size of a minority of galaxies. \adr{SK: The last statement we can actually check, correct? RM: Would need to look at the MSE distribution for spin individually}
% https://ui.adsabs.harvard.edu/abs/2019MNRAS.488.4801J/abstract

\subsection{Nature vs Nurture}
\label{subsec:nature_vs_nurture}
We here define the nature vs nurture problem as the question whether the properties of a galaxy can be determined if you know its state at a single point in time, or if one needs to consider its evolution through time. Feature importance is a useful approach to this question as it allows us to distinguish whether single points during the evolution of galaxies have a large impact on their present-day properties. 
By considering a wide range of snapshots as inputs our method provides insight into this question in a way previous approaches could not. Our results suggest that nurture is more important. If nature was the most important we would expect the feature importance plots to peak at high redshifts which correspond to the initial conditions of the subhalo. However this is never the case, and for most output features the feature importance goes to zero at very high redshifts. It might be thought that this is because some subhalos cannot be tracked back to $z=7.6$ and so are skewing the peak of the feature importance plots to lower redshift, but if we calculate feature importance plots by training models using only subhalos that can be tracked to $z=7.6$ we still do not find a peak at that point. Instead we find a peak at later points, which we discuss in section \ref{subsec:best_snapshots}. Even for properties whose feature importance peaks at early times, such as stellar mass, the feature importance is still high around $z=0$. This shows that the evolution of the host halo at late times always plays a key role in determining the redshift zero galaxy properties.

% \adb{One might argue that the galaxies properties are not set by it's initial conditions, but are instead set a a later snapshot, likely the one at which the feature importance peaks. To investigate this we look at the sum of the feature importance of all the input properties at each snapshot. In Table \ref{table:feature_importance_peak} we show the maximum value obtained for each for the output features. For all output features being predicted the highest feature importance for a single snapshot is less than 20\%. This shows that the majority of the predictive power of the model cannot come from a single point in time. Therefore the physical processes that determine the value of the output property cannot occur at a single point in time. SFR has the highest fraction of it's feature importance in its most important snapshot. This indicates that of the output features we consider it is the most set by nature rather than nurture.}

To quantify nature vs nurture we consider the following integrals of the feature importance over time,
\begin{equation}
    \label{eq:int_1}
    I_{\rm{nat}} = \int_{t_{peak}-\frac{\Delta t}{2}}^{t_{peak}+\frac{\Delta t}{2}} F(t) dt
\end{equation}
\begin{equation}
    \label{eq:int_2}
    I_{\rm{nur}} = \int_{t_0}^{t_{peak}-\frac{\Delta t}{2}} F(t) dt
    + \int_{t_{peak}+\frac{\Delta t}{2}}^{t_f} F(t) dt
\end{equation}
where $t_0$ is the earliest time considered, $t_f$ is the final time considered, $ t_{peak}$ is the time at which the feature importance peaks,  $\Delta t$ is the time around the peak to consider, and $F(t)$ is the feature importance over time. 

Although the values of feature importance we obtain are at single points in time, we argue that feature importance can be treated as a continuous quantity, and therefore integration is a valid technique. This is because the physical properties that are used as input features are well-behaved, i.e. they evolve smoothly and do not exhibit any large discontinuities. We show in Appendix \ref{sec:appendix_int_fi} that if you consider a smaller spacing of snapshots then the trends shown in Figure \ref{fig:feature_importance} are unaffected. In Appendix \ref{sec:appendix_int_fi} we also create a toy model for which we know the relative importance of nature vs nurture. We investigate the integration of the feature importance plots from the toy model, showing that the integrals in equations \ref{eq:int_1} \& \ref{eq:int_2} indeed allow us to distinguish between nature vs nurture.

\begin{figure}   
    \centering
    \includegraphics[width=.47\textwidth]{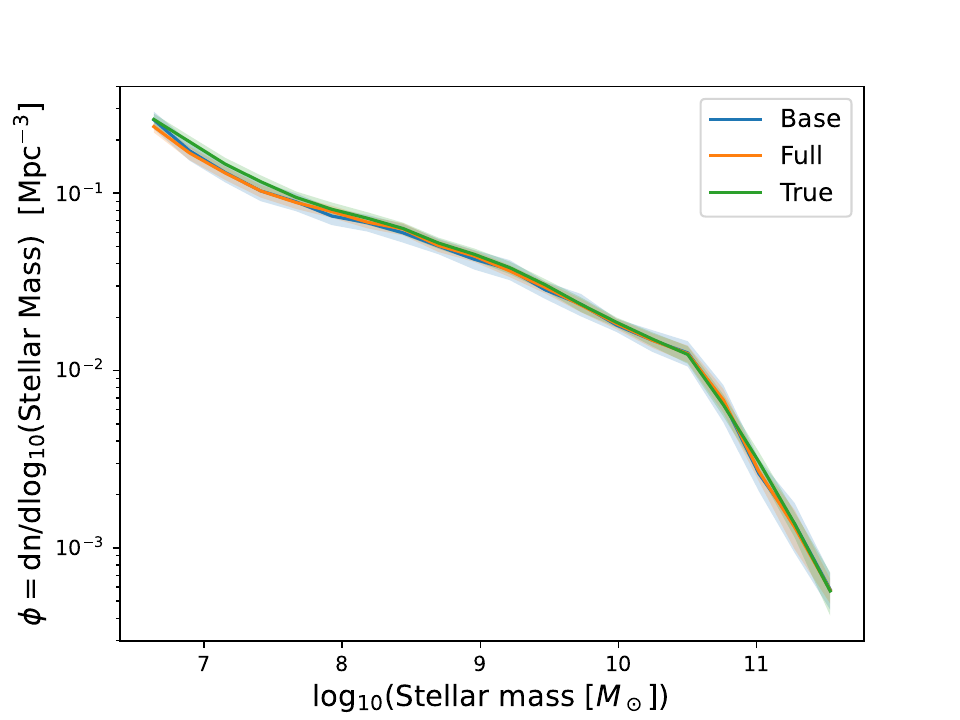}
    
    \caption{The stellar mass function of 10 different test sets. The green line shows the true values, taken directly from the IllustrisTNG simulation, and the green shaded area represents one standard deviation in mass function values from the 10 test/train splits. The blue line shows the predicted stellar mass function from the baseline model, and the orange line shows the prediction from the ERT model that takes in the halo properties from 10 snapshots, starting at redshift $z=7.6$.}
    \label{fig:stellar_mass_function}
\end{figure}

To evaluate the integrals in equations \ref{eq:int_1} and \ref{eq:int_2} we need to choose a value for $\Delta t$. A natural choice would be the dynamical timescale, as any environmental effects associated with the galaxy evolving by nurture will take longer than this to affect the galaxy.
We calculate the dynamical timescale for all subhalos in our sample with
\begin{equation}
    t_{dyn} = \left(\frac{2R^3}{GM}\right)^{\frac{1}{2}} 
\end{equation}
where $M$ is the subhalo dark matter mass and $R$ is twice the radius of the rotation curve maximum. We set $\Delta t=1.5$Gyr, as we find that 99\% of subhalos have a dynamical time less than this. We calculate the ratio of $I_{\rm{nat}}$ to $I_{\rm{nur}}$ for each of the output properties, and show the results in Table \ref{table:feature_importance_peak}.
For properties that peak at $z=0$, we integrate between $t_f - \Delta t$ and $t_f$.
For all output features being predicted $I_{\rm{nur}}$ is significantly larger than $I_{\rm{nat}}$. This shows that the majority of the predictive power of the model cannot come from a single snapshot. Therefore the physical processes that determine the value of the output property cannot occur at a single point in time. SFR has the highest fraction of it's feature importance in its most important snapshot. This indicates that of the output features we consider it is the most set by nature rather than nurture.

% If we consider the dynamical timescale for FOF halos, it is given by
% \begin{equation}
    % t_{dyn} = \left(\frac{2R^3}{GM}\right)^{\frac{1}{2}} 
    % = (100G\rho_{mean}(z))^{-\frac{1}{2}}
% \end{equation}
% where we have used the common definition of the halo boundary that defines a volume with mean density that is 200 times the mean density of the Universe, at the time the halo is considered. $t_{dyn}$ will be maximised at $z=0$, as $\rho_{mean}$ will be minimized at this point. Therefore we end up with $t_{dyn} <= 4.2 $Gyr

% See the arxiv meeting paper - https://arxiv.org/pdf/2101.08549.pdf

\subsection{Stellar mass function}
\label{subsec:mass_function}

In Figure \ref{fig:stellar_mass_function} we show the stellar mass function of the subhalos in the test set. We compare the mass function from the true values compared with the baseline model and the model that takes in the halos full history properties. The shaded area represents 1 standard deviation from 10 different random choices of the train-test split. Both the baseline and full models agree with the true values in that they are within one standard deviation over the full range of stellar masses considered. This indicates that if the only aim of populating a dark matter only simulation is to reproduce mass functions, then using the baseline model is sufficient. However, as is shown in Table \ref{table:model_performance}, including halo history leads to improved performance for individual galaxies. Therefore, future work in this area must focus on further improving quantifiable metrics such as the MSE, as opposed to stopping when mass functions have been matched.

\subsection{Best snapshots to use for predictions}
\label{subsec:best_snapshots}

\begin{figure}   
    \centering
    \includegraphics[width=.47\textwidth]{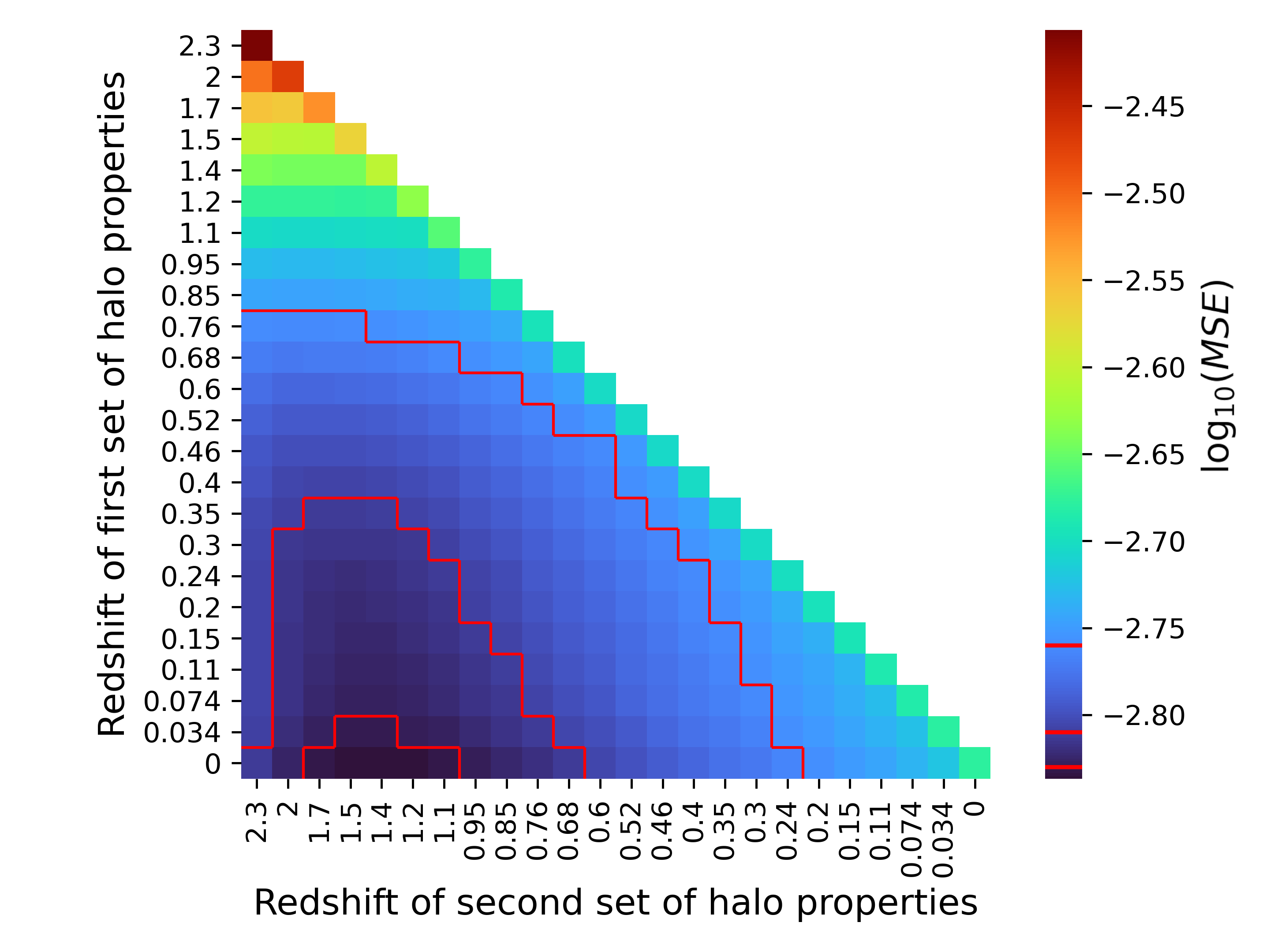}
    
    \caption{Heatmap showing MSE scores when predicting subhalo stellar mass at $z=0$ when halo properties from two snapshots are used as input to an ERT model. MSE scores have been logged to better highlight trends. Red lines indicate contours of constant MSE. The diagonal corresponds to input features from a single snapshot only, i.e. our baseline model. Adding a second snapshot gives significantly improved results, with the best performance coming when the first and second snapshot are well spaced.}
    \label{fig:min_snapshot_heatmap}
\end{figure}

We wish to further verify that the trends shown in the feature importance plots are physical. To do this we train a number of models using halo properties from two different snapshots to predict the stellar mass at redshift zero. In Figure \ref{fig:min_snapshot_heatmap} we show the MSE scores that result from these different models. To generate these scores we used the ERT algorithm, but we have verified that the plot looks similar when other machine learning algorithms are used. The red lines have been added to highlight trends, and indicate contours of constant MSE.

The diagonal of Figure \ref{fig:min_snapshot_heatmap}, where the first and second snapshot are equal, corresponds to a model trained on a single snapshot. Looking at the trend along this diagonal we can see that predictions are worst when using halo properties from a high redshift. As the redshift of the input properties decreases the model performance increases, but the increase plateaus around $z=0.6$. When generating similar figures for predicting the other baryonic properties, this plateau happens at different redshifts. The plateau occurs latest for SFR and gas mass, agreeing with where their peak in feature importance appears.

When looking at the off-diagonal elements which correspond to models with input features taken from two snapshots, we see significant improvements compared with the single snapshot case. This highlights the fact that even adding one more snapshot already has a strong impact, and that by limiting inputs to a single snapshot we significantly hinder the ability of the model to make predictions.  
% he best MSE value is achieved when using input features from the snapshots at $z=0$ and $z=1.4$. 
We would expect this behaviour for any system for which nurture is more important than nature.
It reflects the underlying physics behind determining galaxy properties which is that it is key to consider as much information as possible about their history/evolution. 
% \adr{SK: good. Here is the main key to the multi-epoch ML, since nurture is important multi-epoch ML is needed. If it would be only nature, than we could get away with one snapshot since it would be possible to link in some way the initial conditions to the conditions at that one snapshot. So basically this message must become clear: any system for which nurture is more important than nature does better with multi-epoch ML. }
% Shape of contour lines - Not sure what to say
% Note that 0.6 + 0.52 is better than 0 + 0.2

\begin{figure}   
    \centering
    \includegraphics[width=.47\textwidth]{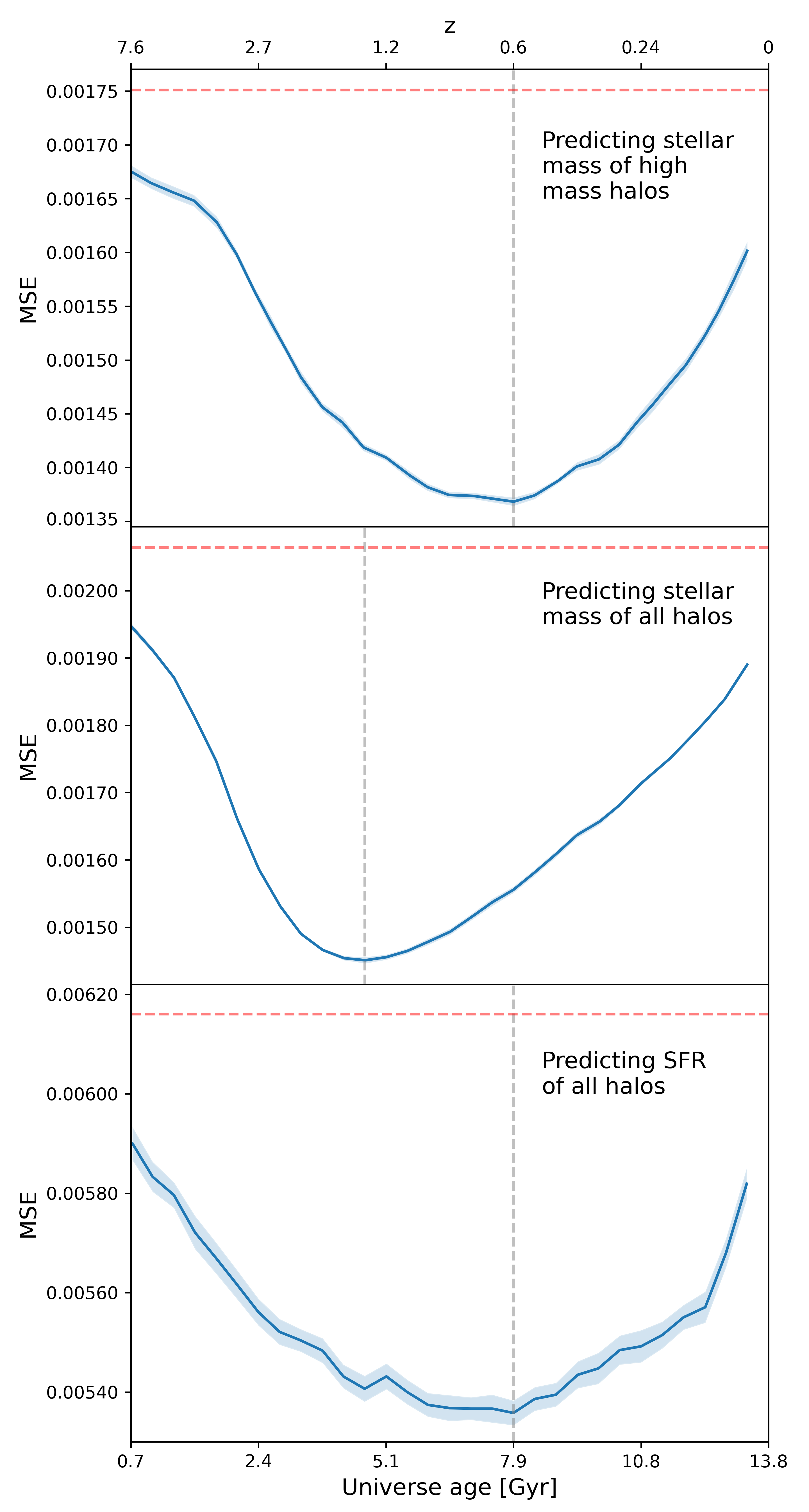}
    
    \caption{The effect of varying the second snapshot of halo properties passed to our models. The first snapshot halo properties are fixed to be redshift 0. The shaded region represents one standard error from 10 train/test splits. The minimum MSE score is shown by the grey dashed line. The performance of the baseline model is indicated by the red dashed line. (\textbf{Top}) Prediction of stellar mass for halos with $M_{DM} > 10^{11}$ (\textbf{Middle}) Prediction of stellar mass for all halos (\textbf{Bottom}) Prediction of SFR for all halos}
    \label{fig:second_snap}
\end{figure}

In Figure \ref{fig:second_snap} we train models to predict redshift zero baryonic properties. We always pass the $z=0$ dark matter properties as input, and we vary the redshift of the second set of halo properties fed into the model. Thus the central plot, which predicts the stellar mass of the entire subhalo population, is equivalent to the the bottom row of Figure \ref{fig:min_snapshot_heatmap}. The red dashed line shows the MSE score of the baseline model. The grey dashed line shows the minimum point. In the bottom panel we show the prediction for the SFR rate. As the $z=0$ properties are already provided as the first snapshot, the minimum of the second snapshot does not occur at redshift zero. It is significantly later than the minimum in the middle panel, again confirming that the different locations of the peaks in the feature importance plots are not just artefact of the ERT algorithm. In the top panel we predict the stellar mass of subhalos which have $M_{\rm{DM}} > 10^{11}$. The minimum occurs much later than in the middle panel. This shows evidence of hierarchical formation. For high mass subhalos a large fraction of their stellar mass comes from mergers with other subhalos. As we only consider the main progenitor branch, when considering a high redshift snapshot much information about the assembly history of the halo will not be included. For low mass subhalos which have evolved without many interactions, their growth history is smoother, and so taking an early snapshot does not lead to information being missed out. As the halo mass function is biased to low mass halos, the minimum for the MSE occurs at early times. If the hierarchical growth model was not correct we would expect the location of the minimums to be reversed, as large halos would form earlier than their smaller counterparts.

% Plots of correlations between output features
% Include more input features over redshifts

\section{Discussions and Conclusions}
\label{sec:discussions}

Figure \ref{fig:second_snap} shows that when looking at subsamples of the subhalo population the MSE scores will differ (as shown by the value of the red dashed line in the top vs middle panel), as will the most important snapshot (as shown by the grey dashed line). This suggests that if we look at the feature importance for models trained on different subsamples of the population we should get different feature importance plots. We find this to be the case.
This shows how our model can pick out the fact that different populations of galaxies form in different ways. In general the relative importance of each halo property remains the same, but the peak moves, indicating that different times are most important for the formation of galaxy subpopulations. However, in some cases we find that the relative importance of the feature importance changes, indicating a different formation or evolution mechanism. We defer a detailed look into the feature importance plots of subsamples of galaxies until a later work.

In Appendix \ref{sec:appendix_single_snapshot} we look at the feature importance from the baseline model when predicting stellar mass at redshifts other than zero. The relative importance of each property agrees between the full model feature importance for stellar mass shown in Figure \ref{fig:feature_importance}. The halo velocity dispersion is most important property at the highest redshifts with halo maximum velocity becoming the most important at later times. Spin is always determined to be the least important. Comparing the feature importance between the full model and single snapshot predictions yields similar results for the other baryonic properties. This is another reassuring test of the robustness of the multi-epoch approach. The halo properties that determined the stellar mass at $z=1$ should remain the most important when looking at the relative importance of $z=1$ input features of the full model. We expect this because most stars present in the galaxy at $z=1$ remain in the galaxy at $z=0$.

% Cumulative feature importance
% Large vs small boxes
% Plots for snapshot range of other output features
% Subsample central vs satellite

\subsection{Conclusions and future work}
\label{subsec:conclusions}

Our conclusions can be summarized as follows.
\begin{itemize}
    \item We introduce a novel method of predicting the baryonic properties of subhalos from dark matter only simulations using machine learning. Our model takes subhalo properties from a wide range of redshifts as input, and can be trained on any simulation with merger trees available. 
    \item When compared with a baseline model that only uses $z=0$ input features, the new model yields significantly more accurate predictions. It also outperforms a model which only uses the mass history of subhalos. Therefore future work which predicts baryonic properties should include a variety of subhalo properties taken over a range of redshifts as their input.
    \item We use a normalized version of mean squared error as our loss function, which allows us to determine which output properties are most difficult to predict.
    \item Using decision tree based algorithms allows us to determine the relative importance of each input feature. Our main plot, Figure \ref{fig:feature_importance}, shows how the feature importance varies depending on the output feature being predicted. This allows us to infer information about how the different baryonic properties of a subhalo are determined, especially the redshift which is most important.
    \item Our feature importance plots and ratios of $I_{\rm{nur}}/I_{\rm{nat}}$ show that for the IllustrisTNG simulations nurture is more important than nature in determining the properties of a galaxy
    \item We confirm our feature importance are not an artefact of the ERT algorithm used in the work by examining how the MSE varies depending on what snapshots are passed to the model. These results hold for machine learning algorithms not based on decision trees.
\end{itemize}

In future work we aim to train the method on other hydrodynamical simulations. It is difficult to untangle the impact of the various subgrid models used by simulations, but the feature importance plots shown in this work may be able to pick out the differences in a quantifiable manner. 

Instead of using dark matter only properties as input, we plan to pass baryonic properties as input. Although the models produced will not be useful for populating dark matter only simulations, their feature importance should give information about how the values of one baryonic property influences the evolution of other properties. We will also include more information in the model than just the main progenitor branch from the merger trees.

\section{Acknowledgements}
The authors are grateful to the anonymous referee for their useful suggestions and questions that helped improved this publication.
RM acknowledges support from an STFC Studentship (Ref. 2145045)

\subsection{Software Citations}
The analysis in this paper  depended  on  the following packages of the python programming language NumPy \citep{numpy}, Matplotlib \citep{matplotlib}, and Scikit-learn \citep{sklearn}; We are thankful to the developers of these tools. This research has made intensive use of NASA’s Astrophysics Data System (http://ui.adsabs.harvard.edu) and the arXiv eprint service (http://arxiv.org).

\section{Data availability}
All simulation data used in this work is available from the IllustrisTNG website (http://www.tng-project.org).
The code used to produce this paper will be provided upon reasonable request.



\bibliographystyle{mnras}
% \bibliography{bib/illustris, bib/introduction, bib/ml, bib/python, bib/similar_work, bib/results}
\bibliography{mnras_template.bib}

%\begin{thebibliography}{99}
%\end{thebibliography}

%%%%%%%%%%%%%%%%% APPENDICES %%%%%%%%%%%%%%%%%%%%%

\appendix

% https://tex.stackexchange.com/a/406922
\renewcommand{\thefigure}{A1}
\begin{figure*}
    \centering
    \includegraphics[width=.38\textwidth]{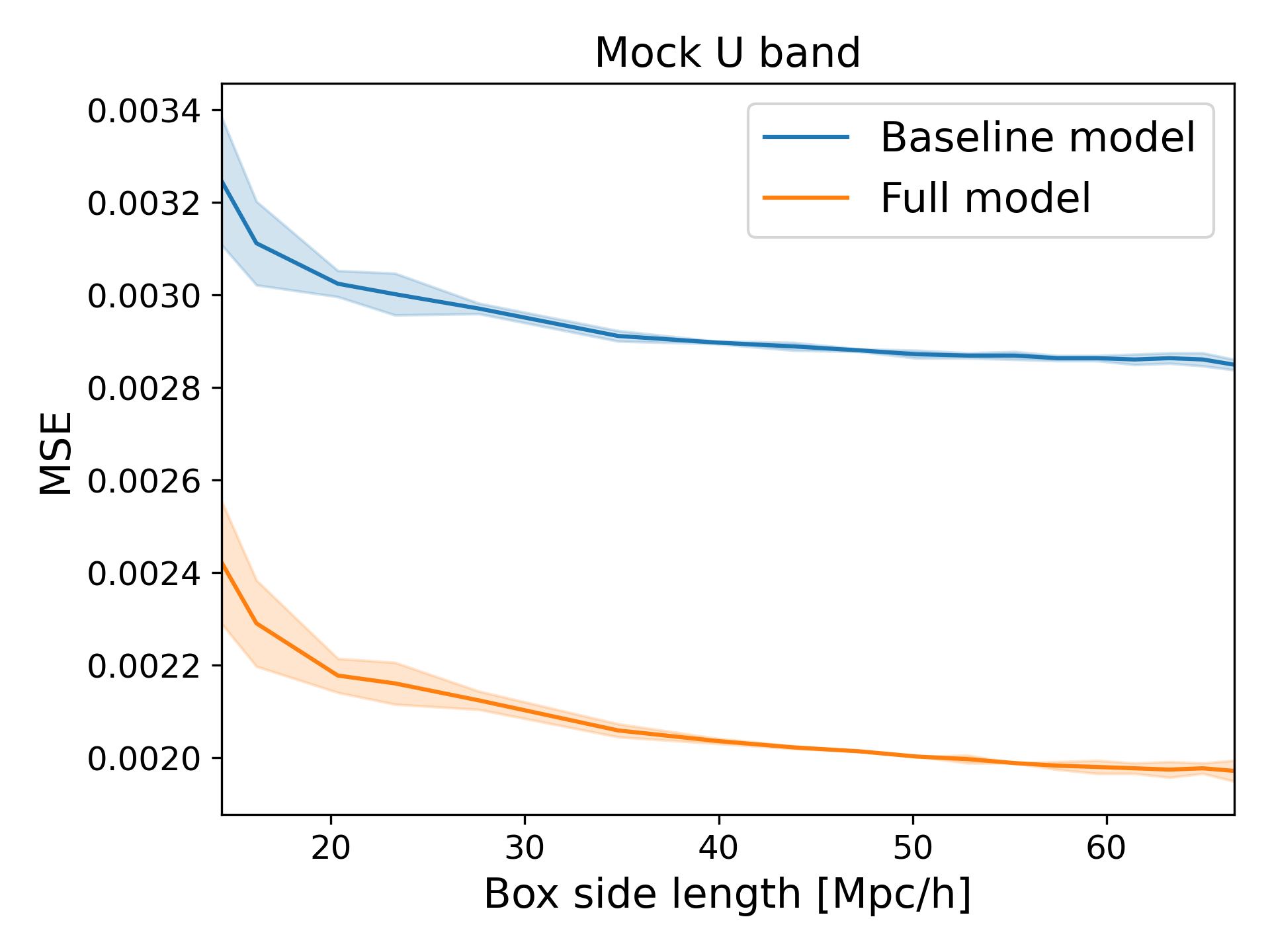}
    \hspace{1cm}
    \includegraphics[width=.38\textwidth]{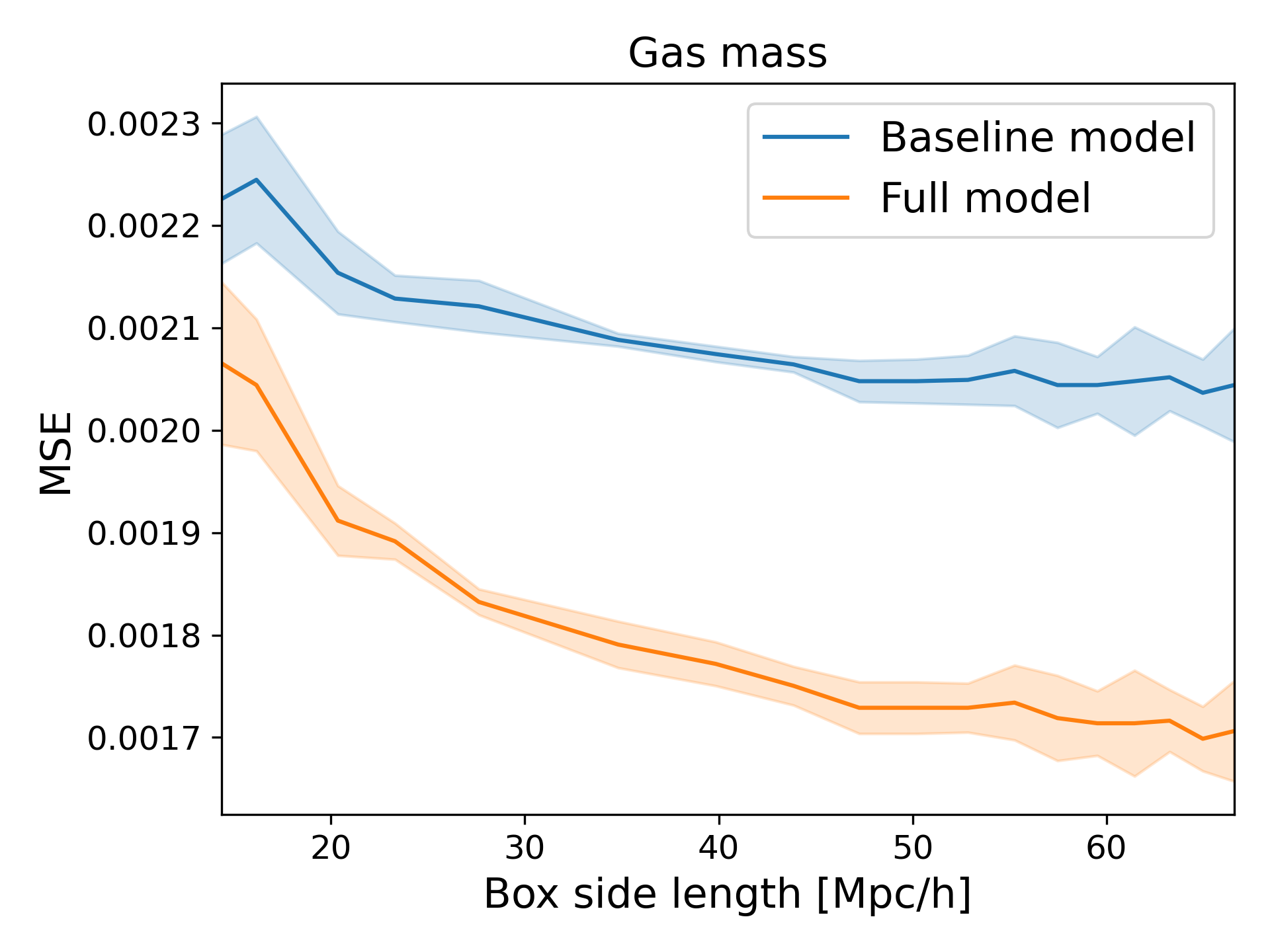}
    
    \vspace{5mm}
    
    \includegraphics[width=.38\textwidth]{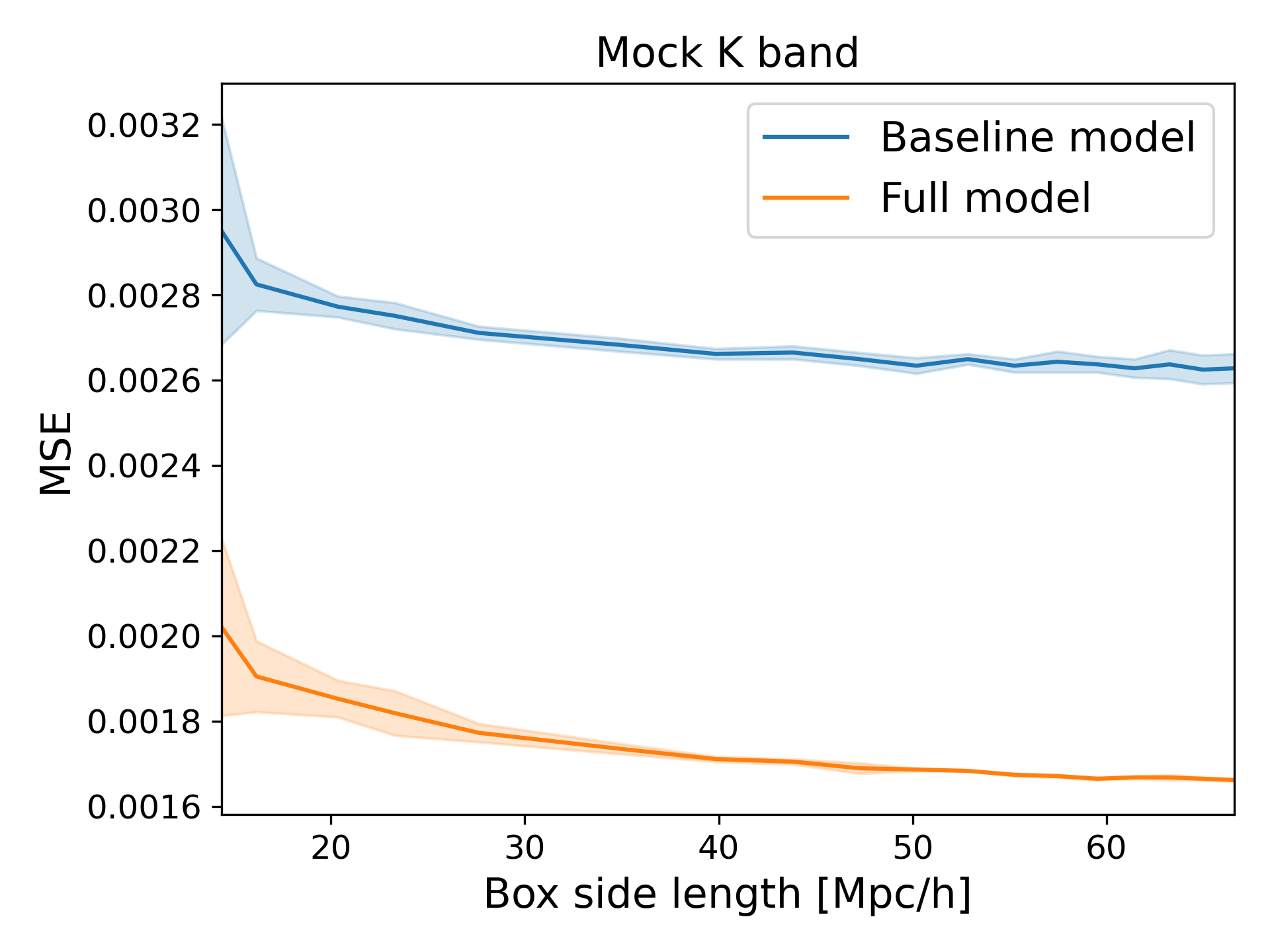}
    \hspace{1cm}
    \includegraphics[width=.38\textwidth]{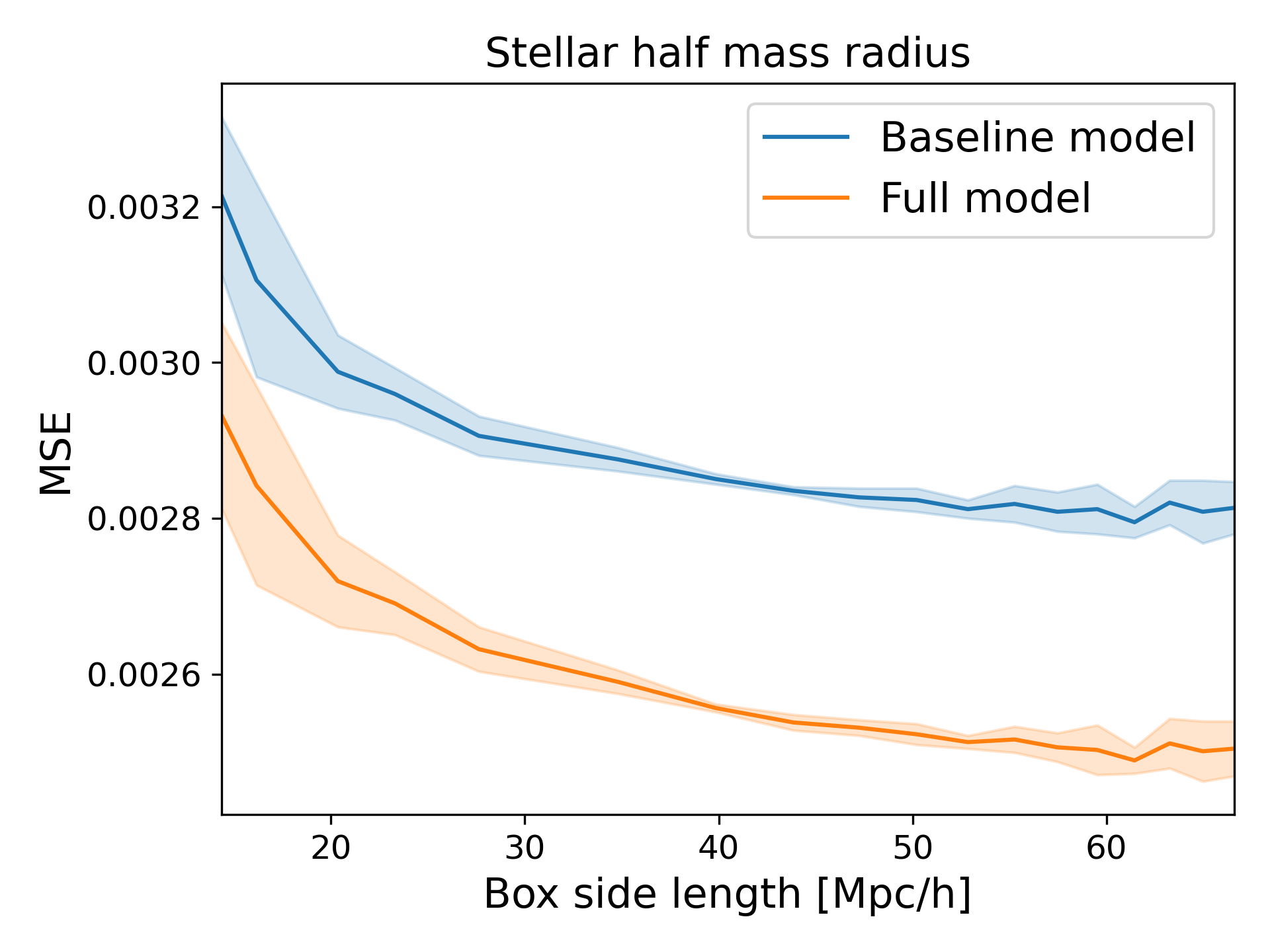}
    
    \vspace{5mm}
    
    \includegraphics[width=.38\textwidth]{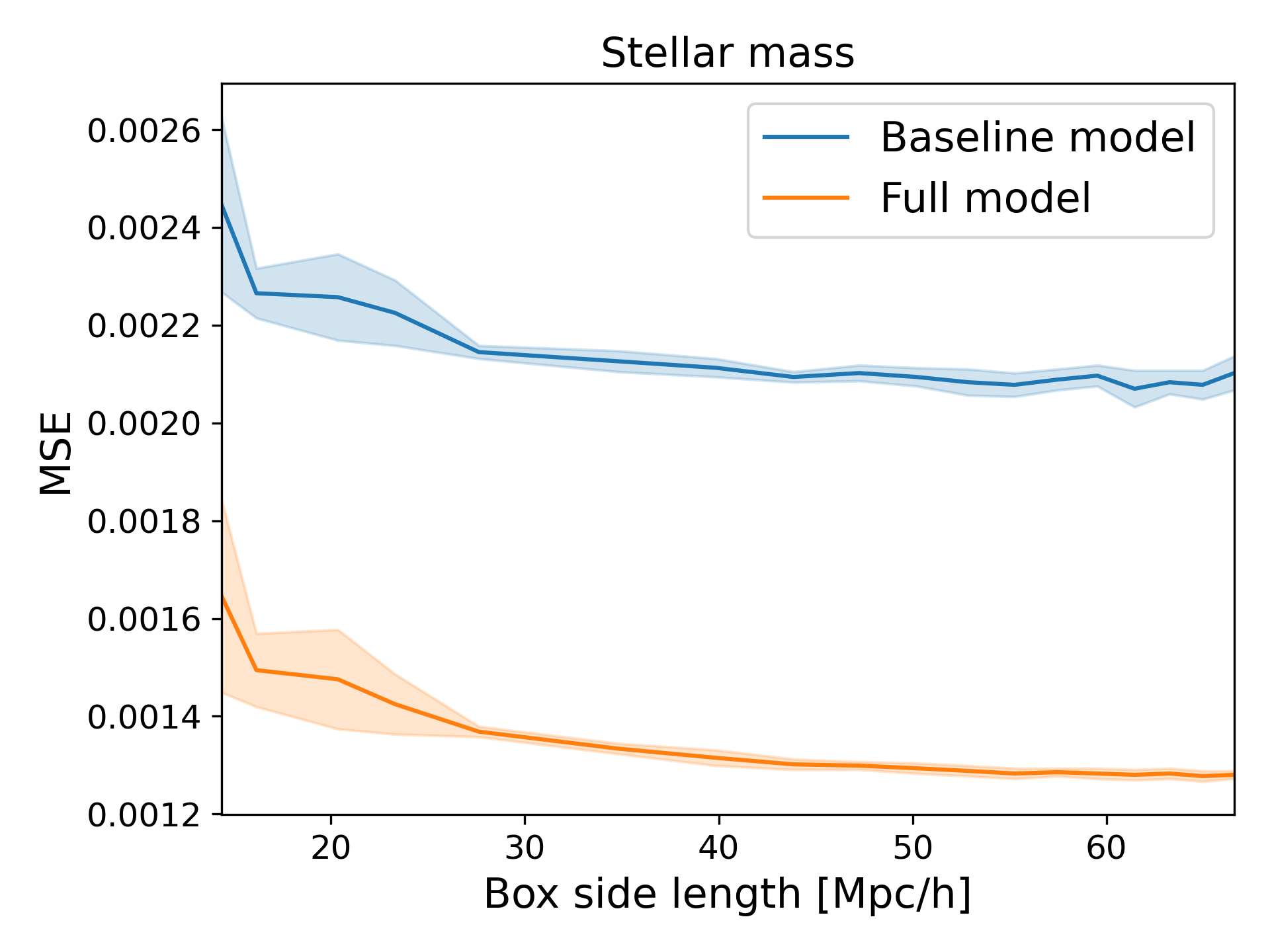}
    \hspace{1cm}
    \includegraphics[width=.38\textwidth]{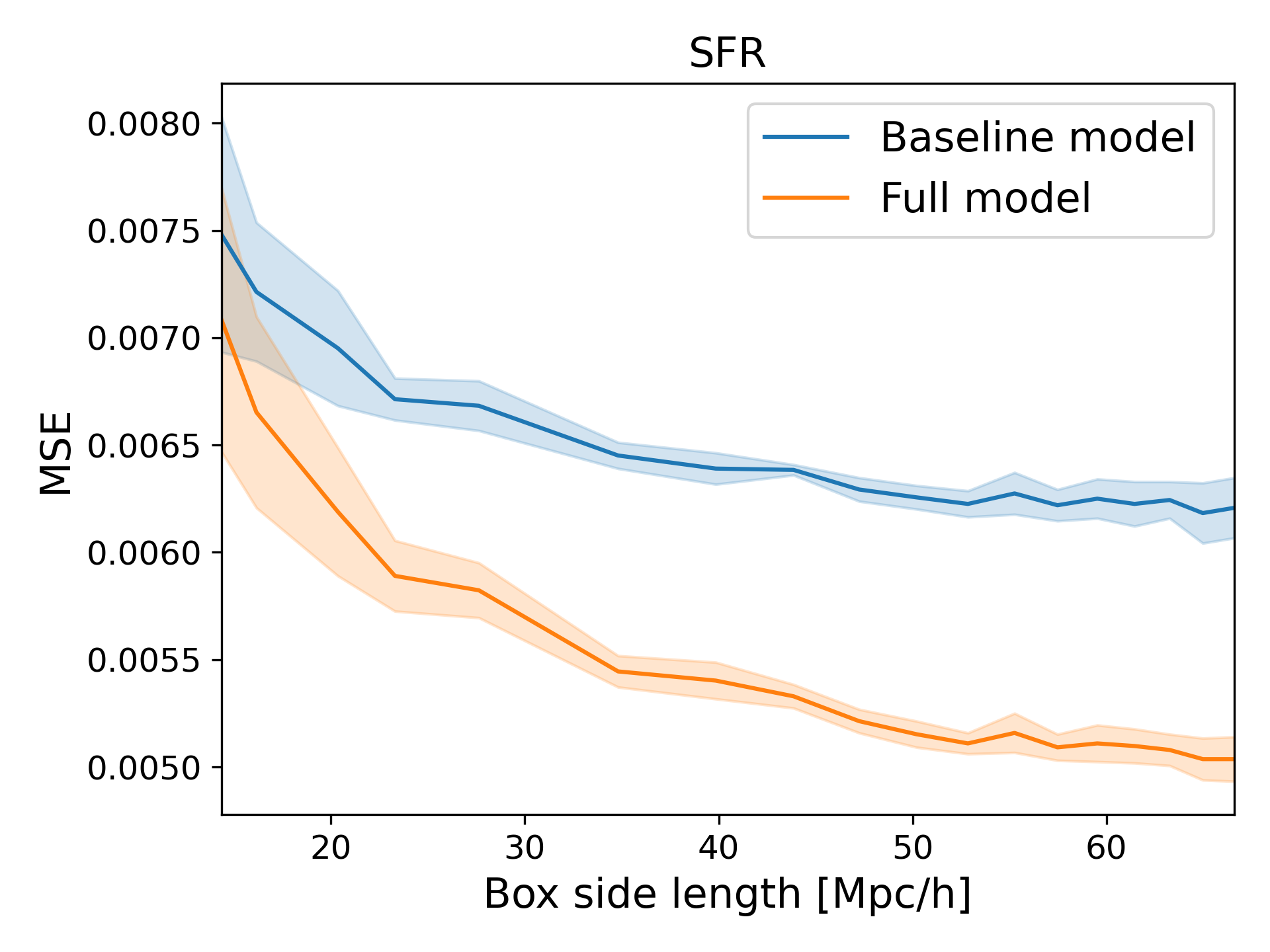}
    
    \vspace{5mm}
    
    \includegraphics[width=.38\textwidth]{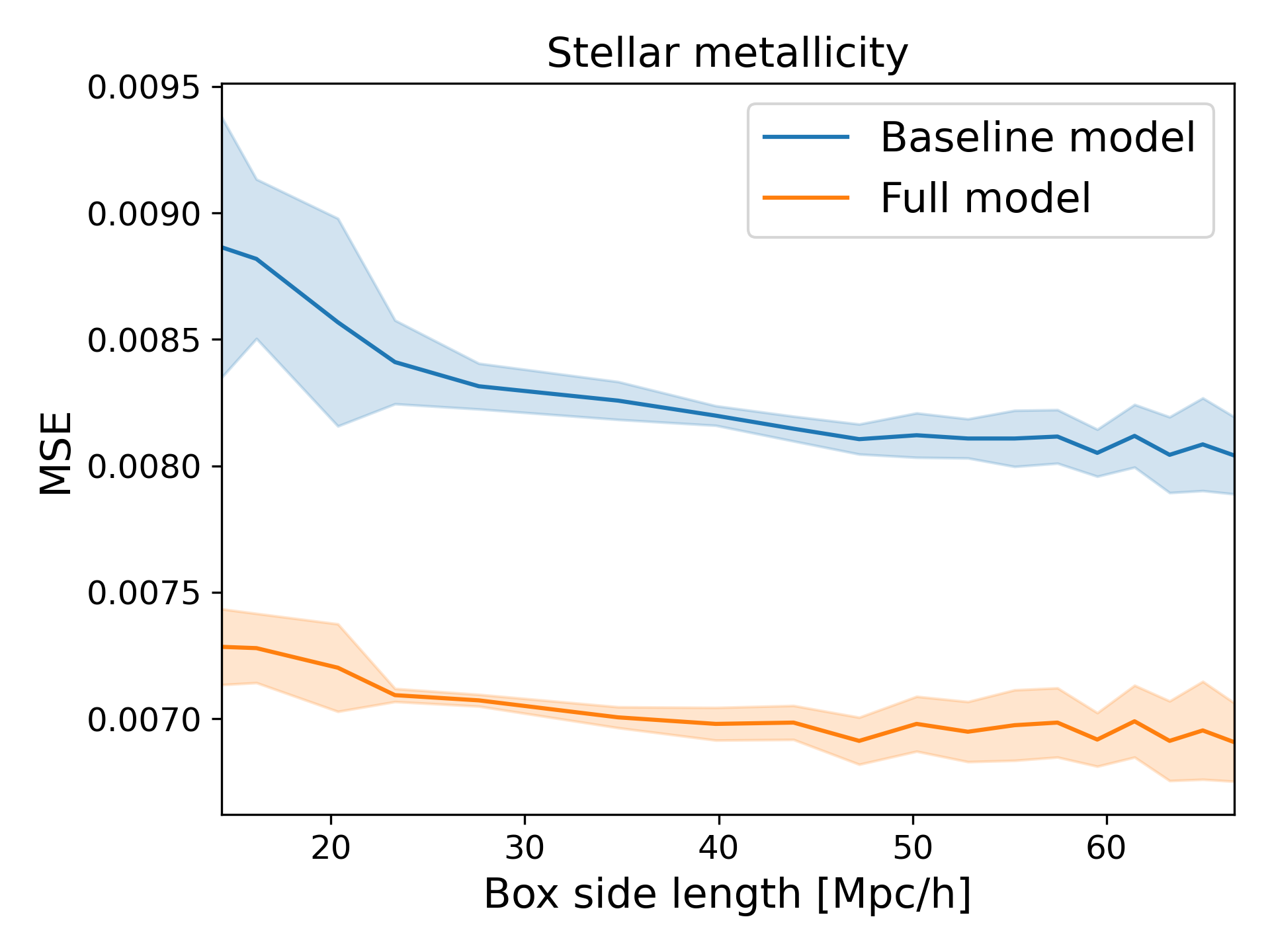}
    \hspace{1cm}
    \includegraphics[width=.38\textwidth]{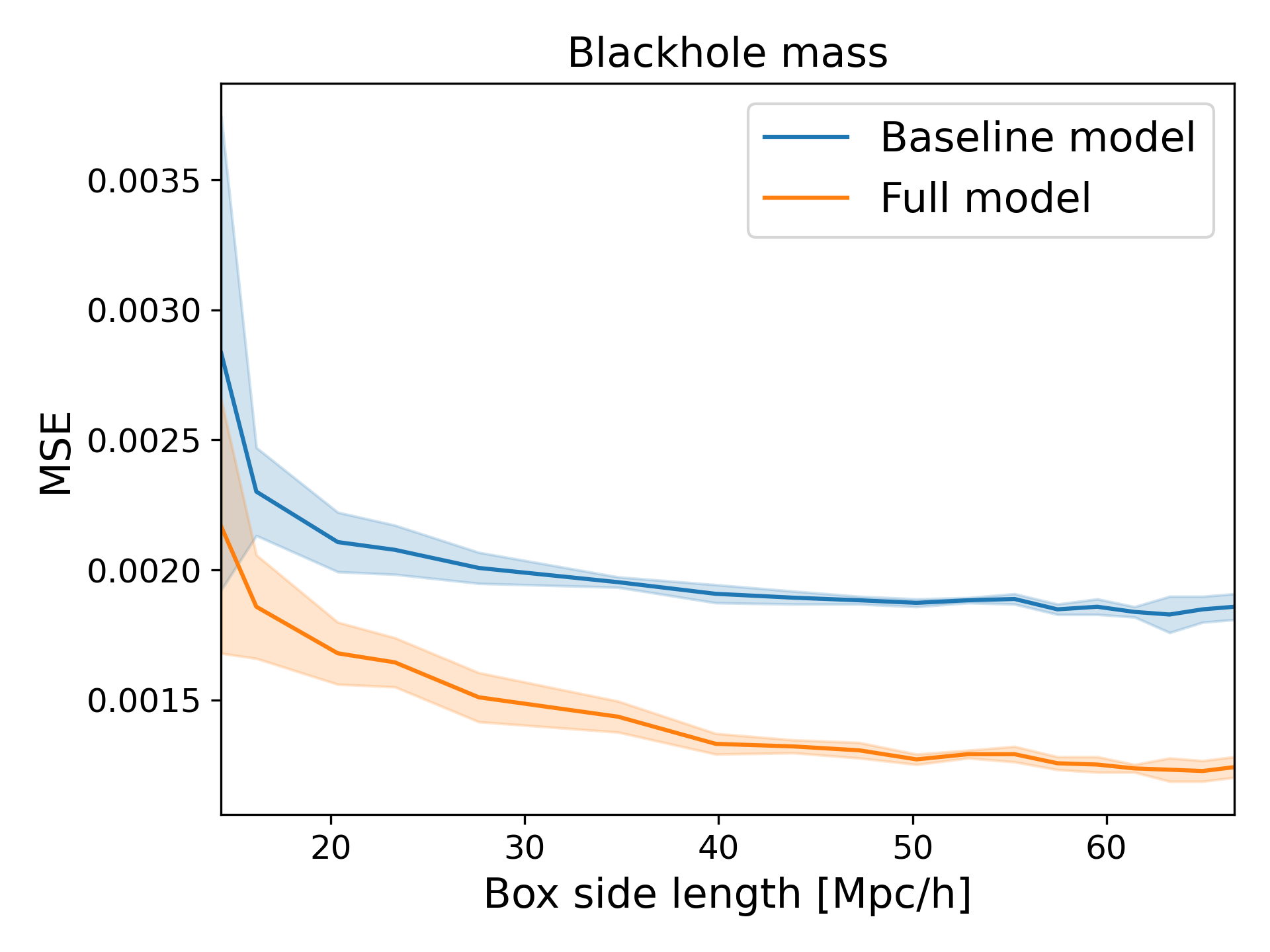}

    \caption{Effect of the size of the training set box length on the performance of the models. Each plot represents the learning rate for one baryonic output feature. The shaded area represents one standard deviation in values for 10 different train/test split.}
    \label{fig:learning_rate}
\end{figure*}

\section{Learning rate for different models}
\label{sec:appendix_learning_rate}

Figure \ref{fig:learning_rate} shows the learning rates for each model predicting baryonic properties.As expected, the performance of both the baseline and full model improves as the size of the training data increases. The difference between the performance of the two types of models is roughly constant as the training set gets larger, with both plateauing around the same point. 

The shaded region represents one standard deviation in the values from the models being trained on 10 different training sets, and so shows the performance range that can be expected. When the training set is a region of high density the model performance is better. The shaded error decreases initially due to the variability in the size of training set, and increases again for larger training boxes as the size of the test set decreases.
 
The learning rates changes if the minimum mass of subhalos is increased. If the mass cut is high enough there is no difference in performance between the full and baseline models for small box sizes. This is a result of the small size of the training set which means there is not enough data for the machine learning model to pick up on information about formation histories. 

\section{Bayesian optimization and hyperparameters}
\label{sec:appendix_bayes_opt}

Bayesian optimization is useful when finding the arguments $x^*$ that minimize a function, $f(x)$, when $f$ has the following properties: its gradients are not known, it is expensive to evaluate, and its evaluations are noisy. In our case $f$ is given by the MSE of the predictions on the test set by a trained model, and $x$ are the hyperparameters used to train the model. After evaluating $f$ with different values of $x$, we use a Gaussian processes \citep{gaussian_process} to approximate $f$. To decide the next value of $x$ to evaluate, we pick the values that minimize the Lower confidence bound,

\begin{equation}
    LCB(x) = u_{GP}(x) - \kappa \sigma_{GP}(x)
\end{equation}

where $u_{GP}$ is the mean of the fitted Gaussian process, and $\sigma_{GP}$ is its standard deviation. $f$ is then evaluated with the values $x$ that minimize the $LCB$, and the Gaussian process fit is updated with the new information. The value of $\kappa$ sets the exploration-exploitation trade-off. If $\kappa$ is small, then the values of $x$ that minimize the acquisition function will be very close to the minimum of $u_{GP}$. If $\kappa$ is large, the $x$ will be taken from a region with high uncertainty, where $\sigma_{GP}$ is large.

\section{Validating the integration of feature importance plots}
\label{sec:appendix_int_fi}

\renewcommand{\thefigure}{C1}
\begin{figure}   
    \centering
    \includegraphics[width=.47\textwidth]{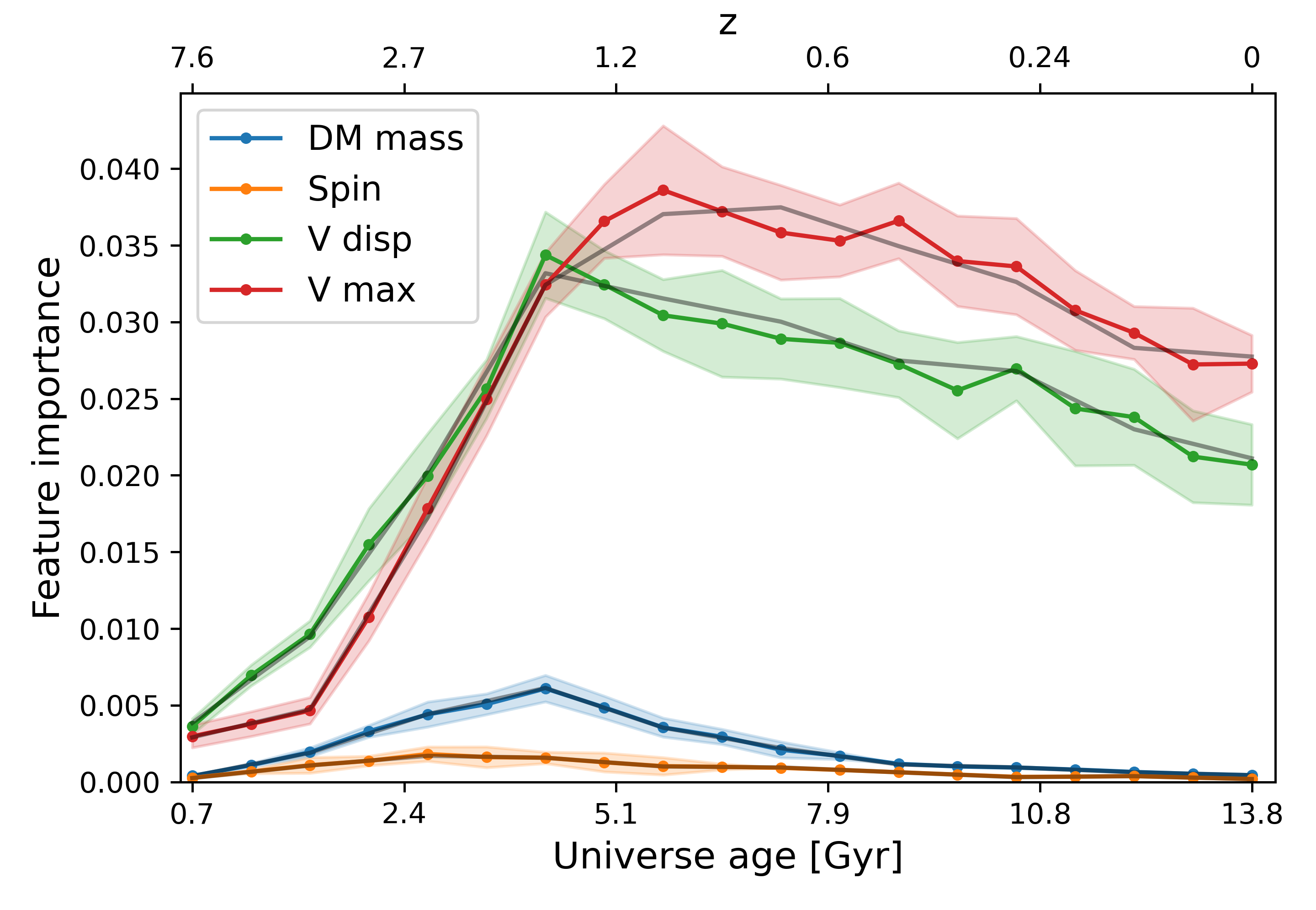}
    
    \caption{
    Feature importance values from the ERT model predicting stellar mass that takes in the halo properties from 19 snapshots starting at redshift $z=7.6$. The grey lines show the feature importance when only using 10 snapshots, as shown in Figure \ref{fig:feature_importance}.}

    \label{fig:spaced_fi}
\end{figure}

For general applications feature importance can be a non-continuous function. However, we here assume that for physical properties which are continuous over time, as is the case for the subhalo properties, the feature importance for that property has to be continuous as well. 
% This is in line with the general implicit assumption in modelling galaxies. 
In Figure \ref{fig:spaced_fi} we show the feature importance values of a model trained using more finely spaced snapshots than Figure \ref{fig:feature_importance}. 
The grey lines show the feature importance values obtained for the original snapshot spacing. As the sum of the feature importance is normalised to one, the grey lines have been rescaled. It should be noted that the standard errors are larger for Figure \ref{fig:spaced_fi} than Figure \ref{fig:feature_importance}.
This comparison of figures allows us to test whether discontinuities could be present in the feature importance that have been smoothed over due to the time step binning. We find that within the limits of the simulation data non such exist, as the grey line always lies within the standard error. This supports our assumption that for this application the feature importance of a single property evolves continuously.

\renewcommand{\thefigure}{C2}
\begin{figure}   
    \centering
    \includegraphics[width=.47\textwidth]{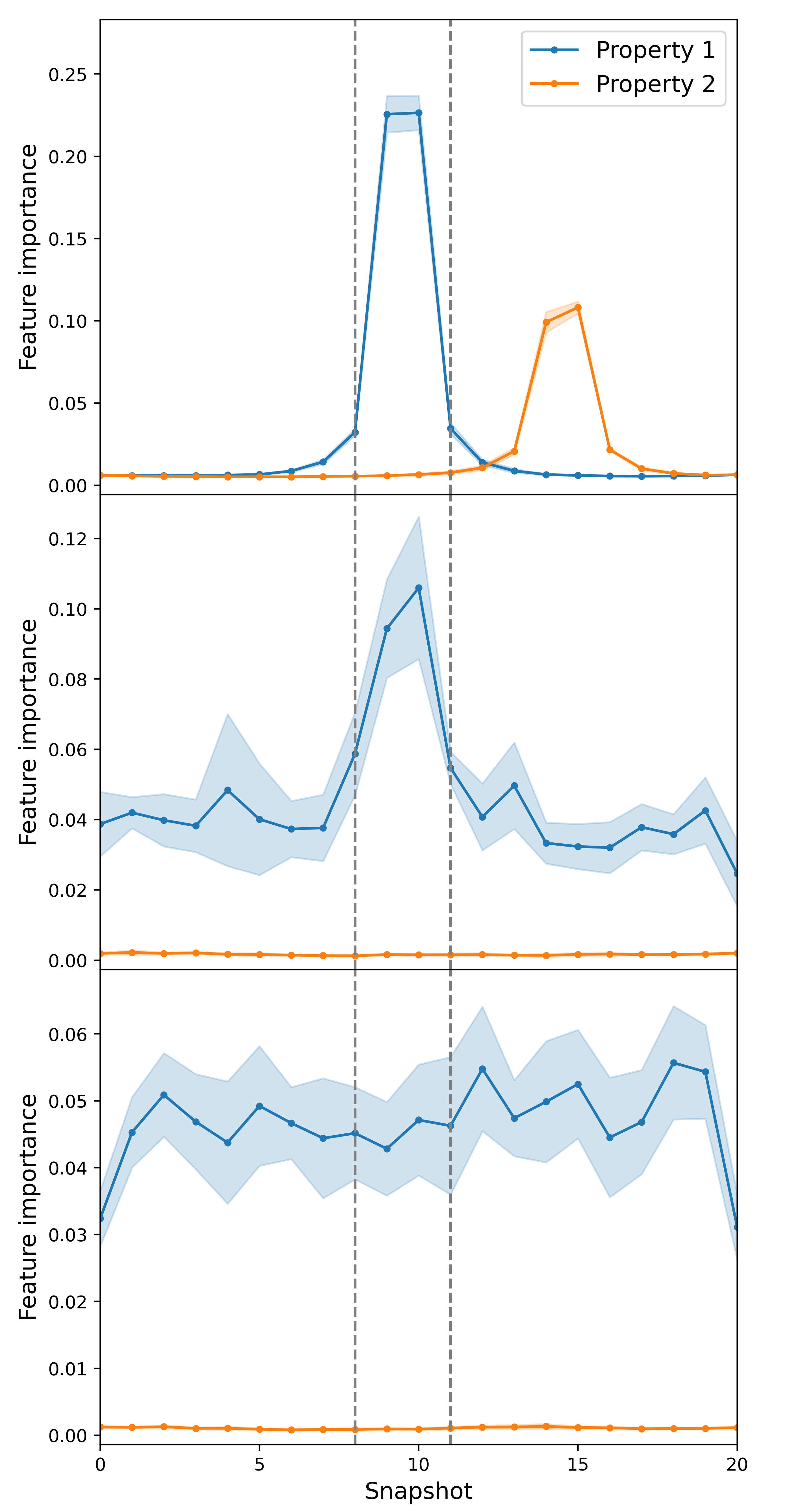}
    
    \caption{
    Feature importance values from an ERT model trained to predict the output features of a toy model. (\textbf{Top}) Nature model (\textbf{Middle}) Mixed model (\textbf{Bottom}) Nurture model
    }
    \label{fig:toy_model}
\end{figure}

To confirm  whether the shapes of the feature importance values in Figure \ref{fig:feature_importance} are due to nature or nurture we consider a toy model. We generate 10000 mock input vectors. For each input vector we generate 2 sets of 21 numbers from $U_{[0,1]}$. For each one of these sets we sort the numbers. A set of 21 number corresponds to a single property sampled over 21 snapshots. This mimics the evolution of the dark matter properties of the halos we consider, which in general grow over time. We then create different output features corresponding to nature and nurture. We train ERT models to predict these output features, and plot their feature importance in Figure \ref{fig:toy_model}. 

For our nature toy model the output feature being predicted is determined by the difference between a property at snapshot $s$ and $s-1$.  In the top panel of Figure \ref{fig:toy_model} the output feature is dependent on the difference between snapshot 10 and 9 from property 1 and snapshots 15 and 14 from property 2. We weight the contribution from property 1 as twice that from property 2. We can see that this results in a feature importance plot with distinct spikes at the snapshots which determine the output feature. Due to our weighting the spike for property 2 is lower than that for property 1. From this top panel we set the limits of our integral as indicated by the grey dashed line.

For our nurture model we the output is given by the sum of the squares of the differences between all consecutive snapshots. In the bottom panel of Figure \ref{fig:toy_model} we show the feature importance of the nurture model. The drop for snapshot 0 and snapshot 20 is because they are only used once in the calculation of the output feature, unlike all other snapshots which are used twice. Within the standard error the feature importance of this model is flat, as we would expect.

For our mixed model we combine the output features of the nurture model and a nature model that only depends on the difference between snapshots 10 and 9 from property 1. We weight the nurture model by a factor of 5. The resulting feature importance is shown in the middle panel of Figure \ref{fig:toy_model}. We can see that the most distinct feature is the spike around snapshot 10, which may initially suggest that nature is more important in the determination of the final output property. However the ratio of the integral within the grey lines to the integral outside the grey lines is equal to 0.35. This is less than the ratio of the weighting, but reflects the fact that the integral for between the grey lines for the pure nurture model is nonzero. Therefore considering the integral of the feature importance plots allows for a comparison of the effects of nature vs nurture.
% \adb{SK: we should again mention that there is a relation between the integral of feature importance and MSE, which we take as our gauge of what is more important when both are present.  }

\section{Feature importance when predicting baryonic properties at higher redshifts}
\label{sec:appendix_single_snapshot}

In Figure \ref{fig:single_snapshot_importance} we look at how the feature importance changes when predicting stellar mass at $z \neq 0$. For this plot we use our baseline model. Note that at each time on the horizontal axis the feature importance will sum to one, whereas in Figure \ref{fig:feature_importance} the sum of feature importance over the whole plot is equal to one. This means there is no peak in feature importance in Figure \ref{fig:single_snapshot_importance}. There are two major effects on the feature importance from using such a small number of input features. First the variation becomes greater. Secondly, the spin feature importance is never zero. This is a result of the model overfitting as it does not have enough features to split on. This can be driven to zero by tuning the \textit{max\_depth} hyperparameter.
 
\renewcommand{\thefigure}{D1}
\begin{figure}   
    \centering
    \includegraphics[width=.47\textwidth]{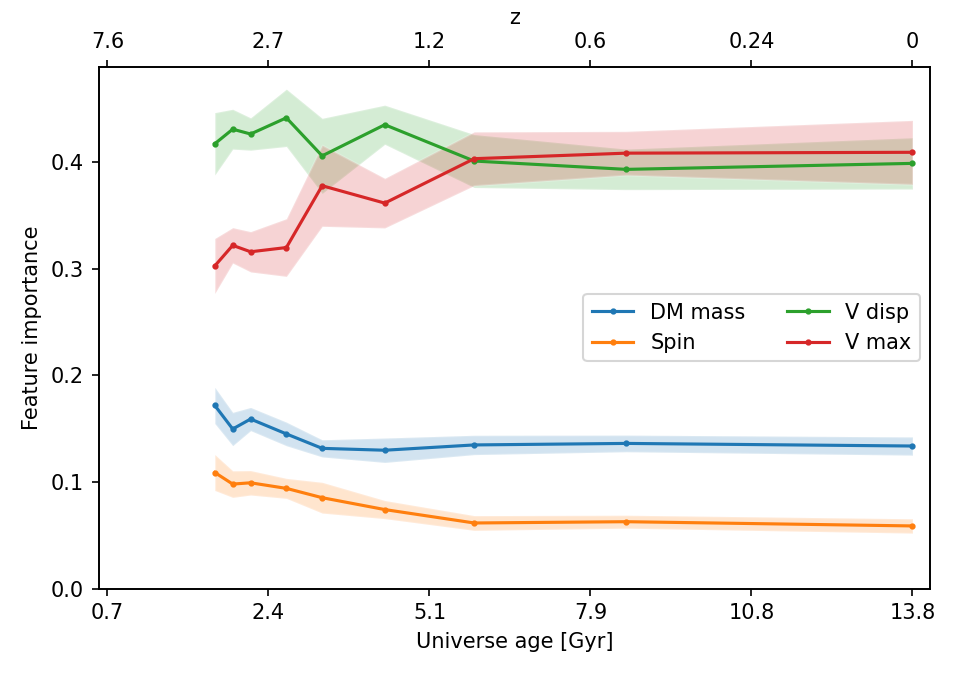}
    
    \caption{The feature importance values from models trained to predict stellar mass at different redshifts. The baseline model is used, with the input being the subhalo dark matter properties from the redshift being predicted.}
    \label{fig:single_snapshot_importance}
\end{figure}

%%%%%%%%%%%%%%%%%%%%%%%%%%%%%%%%%%%%%%%%%%%%%%%%%%

% Don't change these lines
\bsp	% typesetting comment
\label{lastpage}
\end{document}